\documentclass[twocolumn]{aastex63}
\usepackage[utf8]{inputenc}

\usepackage{natbib}
\usepackage{graphicx}
\usepackage{amsmath}
\usepackage{multirow}
\usepackage{booktabs}
\usepackage{longtable}
\usepackage{gensymb}
\usepackage{hyperref}
\usepackage{url}
\usepackage{parskip}
\usepackage{tablefootnote}
\usepackage{gensymb}
\usepackage{wasysym}
\usepackage{float}
\usepackage[T1]{fontenc}
\usepackage[figure,figure*]{hypcap}
\usepackage{lineno}
\newcommand{\totalRVs}{130}
\newcommand{\HIRESRVs}{100}
\newcommand{\HARPSRVs}{28}

\newcommand{\massb}{$\rm{5.3 \pm~1.7}$}
\newcommand{\massc}{$\rm{8.1 \pm~1.1}$}
\newcommand{\massd}{$\rm{8.8 \pm~1.2}$}
\newcommand{\masse}{$\rm{14.8 \pm~2.3}$}

\def\tess{$\it{TESS }$}
\def\kep{$\it{Kepler }$}

\pdfoutput=1 

\begin{document}

\title{The TESS-Keck Survey. XI. Mass Measurements for Four Transiting sub-Neptunes orbiting K dwarf TOI-1246}

\author[0000-0002-1845-2617]{Emma V. Turtelboom}\affiliation{Department of Astronomy, 501 Campbell Hall, University of California, Berkeley, CA 94720, USA}

\author[0000-0002-3725-3058]{Lauren M. Weiss}\affiliation{Department of Physics, University of Notre Dame, Notre Dame, IN, 46556, USA}

\author[0000-0001-8189-0233]{Courtney D. Dressing}\affiliation{Department of Astronomy, 501 Campbell Hall, University of California, Berkeley, CA 94720, USA}

\author{Grzegorz Nowak} \affiliation{Instituto de Astrof\'isica de Canarias (IAC), E-38200 La Laguna, Tenerife, Spain} \affiliation{Departamento de Astrof\'isica, Universidad de La Laguna (ULL), E-38206 La Laguna, Tenerife, Spain} 

\author[0000-0003-0987-1593 ]{Enric Pall\'e} \affiliation{Instituto de Astrof\'isica de Canarias (IAC), E-38200 La Laguna, Tenerife, Spain} \affiliation{Departamento de Astrof\'isica, Universidad de La Laguna (ULL), E-38206 La Laguna, Tenerife, Spain} 

\author[0000-0001-7708-2364]{Corey Beard} \affiliation{Department of Physics \& Astronomy, University of California Irvine, Irvine, CA 92697, USA}

\author{Sarah Blunt} \affiliation{Department of Astronomy, California Institute of Technology, Pasadena, CA 91125, USA}

\author[0000-0002-4480-310X]{Casey Brinkman} \affiliation{Institute for Astronomy, University of Hawai`i, 2680 Woodlawn Drive, Honolulu, HI 96822, USA}

\author[0000-0003-1125-2564]{Ashley Chontos} \altaffiliation{NSF Graduate Research Fellow} \affiliation{Institute for Astronomy, University of Hawai`i, 2680 Woodlawn Drive, Honolulu, HI 96822, USA}

\author[0000-0002-9879-3904]{Zachary R. Claytor} \affiliation{Institute for Astronomy, University of Hawai`i, 2680 Woodlawn Drive, Honolulu, HI 96822, USA}

\author[0000-0002-8958-0683]{Fei Dai}  \affiliation{Division of Geological and Planetary Sciences, 1200 E California Blvd, Pasadena, CA, 91125, USA}

\author[0000-0002-4297-5506]{Paul A.\ Dalba} \altaffiliation{NSF Astronomy and Astrophysics Postdoctoral Fellow} \affiliation{Department of Astronomy and Astrophysics, University of California, Santa Cruz, CA 95064, USA} \affiliation{Department of Earth and Planetary Sciences, University of California Riverside, 900 University Ave, Riverside, CA 92521, USA}

\author[0000-0002-8965-3969]{Steven Giacalone} \affiliation{Department of Astronomy, 501 Campbell Hall, University of California, Berkeley, CA 94720, USA}

\author{Erica Gonzales} \affiliation{Department of Astronomy and Astrophysics, University of California, Santa Cruz, CA 95064, USA}

\author[0000-0001-5737-1687]{Caleb K. Harada} \altaffiliation{NSF Graduate Research Fellow} \affiliation{Department of Astronomy, 501 Campbell Hall, University of California, Berkeley, CA 94720, USA}

\author[0000-0002-0139-4756]{Michelle L. Hill} \affiliation{Department of Earth and Planetary Sciences, University of California Riverside, 900 University Ave, Riverside, CA 92521, USA}

\author[0000-0002-5034-9476]{Rae Holcomb} \affiliation{Department of Physics \& Astronomy, University of California Irvine, Irvine, CA 92697, USA}

\author[0000-0002-0076-6239]{Judith Korth} \affiliation{Department of Space, Earth and Environment, Astronomy and Plasma Physics, Chalmers University of Technology, 412 96 Gothenburg, Sweden}

\author[0000-0001-8342-7736]{Jack Lubin} \affiliation{Department of Physics \& Astronomy, University of California Irvine, Irvine, CA 92697, USA}

\author{Thomas Masseron} \affiliation{Instituto de Astrof\'isica de Canarias (IAC), E-38200 La Laguna, Tenerife, Spain} \affiliation{Departamento de Astrof\'isica, Universidad de La Laguna (ULL), E-38206 La Laguna, Tenerife, Spain} 

\author{Mason MacDougall} \affiliation{Astronomy Department, 475 Portola Plaza, University of California, Los Angeles, CA 90095, USA} 

\author[0000-0002-7216-2135]{Andrew W. Mayo}\affiliation{Department of Astronomy, 501 Campbell Hall, University of California, Berkeley, CA 94720, USA} \affiliation{Centre for Star and Planet Formation, Natural History Museum of Denmark \& Niels Bohr Institute, University of Copenhagen, \O ster Voldgade 5-7, DK-1350 Copenhagen K., Denmark}

\author[0000-0003-4603-556X]{Teo Mo\v{c}nik}\affiliation{Gemini Observatory/NSF's NOIRLab, 670 N. A'ohoku Place, Hilo, HI 96720, USA}

\author[0000-0001-8898-8284]{Joseph M. Akana Murphy} \altaffiliation{NSF Graduate Research Fellow} \affiliation{Department of Astronomy and Astrophysics, University of California, Santa Cruz, CA 95064, USA}

\author[0000-0001-7047-8681]{Alex S. Polanski} \affiliation{Department of Physics and Astronomy, University of Kansas, Lawrence, KS, USA}

\author[0000-0002-7670-670X]{Malena Rice} \altaffiliation{NSF Graduate Research Fellow}\affiliation{Department of Astronomy, Yale University, New Haven, CT 06511, USA}

\author[0000-0003-3856-3143]{Ryan A. Rubenzahl} \altaffiliation{NSF Graduate Research Fellow} \affiliation{Department of Astronomy, California Institute of Technology, Pasadena, CA 91125, USA}

\author[0000-0003-3623-7280]{Nicholas Scarsdale} \affiliation{Department of Astronomy and Astrophysics, University of California, Santa Cruz, CA 95064, USA}

\author[0000-0002-3481-9052]{Keivan G.\ Stassun} \affiliation{Department of Physics and Astronomy, Vanderbilt University, Nashville, TN 37235, USA}

\author[0000-0003-0298-4667]{Dakotah B. Tyler} \affiliation{Astronomy Department, 475 Portola Plaza, University of California, Los Angeles, CA 90095, USA} 

\author[0000-0002-4290-6826]{Judah Van Zandt} \affiliation{Department of Physics \& Astronomy, University of California Los Angeles, Los Angeles, CA 90095, USA}

\author{Ian J.\ M.\ Crossfield} \affiliation{Department of Physics and Astronomy, University of Kansas, Lawrence, KS, USA}

\author[0000-0003-0047-4241]{Hans J. Deeg} \affiliation{Instituto de Astrof\'isica de Canarias (IAC), E-38200 La Laguna, Tenerife, Spain} \affiliation{Departamento de Astrof\'isica, Universidad de La Laguna (ULL), E-38206 La Laguna, Tenerife, Spain} 

\author[0000-0003-3504-5316]{Benjamin Fulton}\affiliation{NASA Exoplanet Science Institute/Caltech-IPAC, MC 314-6, 1200 E California Blvd, Pasadena, CA 91125, USA}

\author[0000-0001-8627-9628]{Davide Gandolfi} \affiliation{Dipartimento di Fisica, Universit\'a degli Studi di Torino, I-10125, Torino, Italy}

\author[0000-0001-8638-0320]{Andrew W. Howard}\affiliation{Department of Astronomy, California Institute of Technology, Pasadena, CA 91125, USA}

\author{Dan Huber}\affiliation{Institute for Astronomy, University of Hawai`i, 2680 Woodlawn Drive, Honolulu, HI 96822, USA}

\author[0000-0002-0531-1073]{Howard Isaacson}\affiliation{Department of Astronomy, 501 Campbell Hall, University of California, Berkeley, CA 94720, USA} \affiliation{Centre for Astrophysics, University of Southern Queensland, Toowoomba, QLD, Australia}

\author[0000-0002-7084-0529]{Stephen R. Kane} \affiliation{Department of Earth and Planetary Sciences, University of California Riverside, 900 University Ave, Riverside, CA 92521, USA}

\author[0000-0002-9910-6088]{Kristine W. F. Lam} \affiliation{Institute of Planetary Research, German Aerospace Center, 12489 Berlin, Germany}

\author[0000-0002-4671-2957]{Rafael Luque} \affiliation{Instituto de Astrof\'isica de Andaluc\'ia (IAA-CSIC), Glorieta de la Astronom\'ia s/n, 18008 Granada, Spain}

\author[0000-0002-1208-4833]{Eduardo\ L.\ Mart\'in} \affiliation{Instituto de Astrof\'isica de Canarias (IAC), E-38200 La Laguna, Tenerife, Spain} \affiliation{Departamento de Astrof\'isica, Universidad de La Laguna (ULL), E-38206 La Laguna, Tenerife, Spain} \affiliation{Consejo Superior de Investigaciones Cient\'ificas (CSIC), E-28006 Madrid, Spain}

\author[0000-0002-4262-5661]{Giuseppe Morello} \affiliation{Instituto de Astrofísica de Canarias (IAC), 38205 La Laguna, Tenerife, Spain} \affiliation{Departamento de Astrof\'isica, Universidad de La Laguna (ULL), E-38206 La Laguna, Tenerife, Spain} 

\author[0000-0003-2066-8959]{Jaume Orell-Miquel} \affiliation{Instituto de Astrofísica de Canarias (IAC), 38205 La Laguna, Tenerife, Spain} \affiliation{Departamento de Astrof\'isica, Universidad de La Laguna (ULL), E-38206 La Laguna, Tenerife, Spain} 

\author[0000-0003-0967-2893]{Erik A. Petigura} \affiliation{Department of Physics \& Astronomy, University of California Los Angeles, Los Angeles, CA 90095, USA}

\author[0000-0003-0149-9678]{Paul Robertson}\affiliation{Department of Physics \& Astronomy, University of California Irvine, Irvine, CA 92697, USA}

\author[0000-0001-8127-5775]{Arpita Roy}\affiliation{Space Telescope Science Institute, 3700 San Martin Drive, Baltimore, MD 21218, USA} \affiliation{Department of Physics and Astronomy, Johns Hopkins University, 3400 N Charles St, Baltimore, MD 21218, USA}

\author[0000-0001-5542-8870]{Vincent Van Eylen} \affiliation{Mullard Space Science Laboratory, University College London, Holmbury St Mary, Dorking, Surrey RH5 6NT, UK}

\author[0000-0002-2970-0532]{David Baker} \affiliation{Physics Department, Austin College, Sherman, TX 75090, USA}
 
\author[0000-0003-3469-0989]{Alexander A.\ Belinski} \affiliation{Sternberg Astronomical Institute, M.V. Lomonosov Moscow State University, 13, Universitetskij pr., 119234, Moscow, Russia}
 
\author[0000-0001-6637-5401]{Allyson Bieryla}  \affiliation{Center for Astrophysics \textbar \ Harvard \& Smithsonian, 60 Garden Street, Cambridge, MA 02138, USA}

\author[0000-0002-5741-3047]{David R. Ciardi} \affiliation{NASA Exoplanet Science Institute, Caltech/IPAC, Mail Code 100-22, 1200 E. California Blvd., Pasadena, CA 91125, USA}

\author[0000-0001-6588-9574]{Karen A.\ Collins} \affiliation{Center for Astrophysics \textbar \ Harvard \& Smithsonian, 60 Garden Street, Cambridge, MA 02138, USA}

\author{Neil Cutting} \affiliation{Physics Department, Austin College, Sherman, TX 75090, USA}

\author[0000-0002-6042-7351]{Devin J. Della-Rose} \affiliation{Physics and Meteorology Department, 2354 Fairchild Drive, Suite 2A91, U.S. Air Force Academy, CO 80840, USA}

\author{Taylor B. Ellingsen} \affiliation{Department of Physics and Astronomy, George Mason University, 4400 University Drive MSN 3F3, Fairfax, VA 22030, USA}

\author[0000-0001-9800-6248]{E. Furlan} \affiliation{NASA Exoplanet Science Institute, Caltech/IPAC, Mail Code 100-22, 1200 E. California Blvd., Pasadena, CA 91125, USA}

\author[0000-0002-4503-9705]{Tianjun Gan}\affiliation{Department of Astronomy, Tsinghua University, Beijing 100084, China}

\author[0000-0003-2519-6161]{Crystal~L.~Gnilka} \affiliation{NASA Exoplanet Science Institute, Caltech/IPAC, Mail Code 100-22, 1200 E. California Blvd., Pasadena, CA 91125, USA} \affiliation{NASA Ames Research Center, Moffett Field, CA 94035, USA}

\author[0000-0002-4308-2339]{Pere Guerra} \affiliation{Observatori Astronòmic Albanyà, Camí de Bassegoda S/N, Albanyà 17733, Girona, Spain}

\author[0000-0002-2532-2853]{Steve~B.~Howell} \affiliation{NASA Ames Research Center, Moffett Field, CA 94035, USA}

\author[0000-0002-5000-9316]{Mary Jimenez} \affiliation{Department of Physics and Astronomy, George Mason University, 4400 University Drive MSN 3F3, Fairfax, VA 22030, USA}

\author[0000-0001-9911-7388]{David W. Latham} \affiliation{Center for Astrophysics \textbar \ Harvard \& Smithsonian, 60 Garden Street, Cambridge, MA 02138, USA}

\author{Maude Larivière} \affiliation{Department of Physics, 3600 rue University, McGill University, Montréal, Québec, H3A 2T8, Canada} \affiliation{Institute for Research on Exoplanets (\emph{iREx})}

\author[0000-0002-9903-9911]{Kathryn V. Lester} \affiliation{NASA Ames Research Center, Moffett Field, CA 94035, USA}

\author[0000-0003-3742-1987]{Jorge Lillo-Box} \affiliation{Centro de Astrobiolog\'ia (CAB, CSIC-INTA), Depto. de Astrof\'isica, ESAC campus, 28692, Villanueva de la Ca\~nada (Madrid), Spain}

\author{Lindy Luker} \affiliation{Physics Department, Austin College, Sherman, TX 75090, USA}

\author[0000-0002-9312-0073]{Christopher~R.~Mann} \affiliation{Université de Montréal, Montréal, QC H2V 0B3, Canada} \affiliation{Institute for Research on Exoplanets (\emph{iREx})}

\author[0000-0002-8864-1667]{Peter P. Plavchan} \affiliation{Department of Physics and Astronomy, George Mason University, 4400 University Drive, Fairfax, VA 22030, USA}

\author[0000-0003-1713-3208]{Boris Safonov} \affiliation{Sternberg Astronomical Institute, M.V. Lomonosov Moscow State University, 13, Universitetskij pr., 119234, Moscow, Russia}

\author{Brett Skinner} \affiliation{Physics Department, Austin College, Sherman, TX 75090, USA}

\author[0000-0003-0647-6133]{Ivan A. Strakhov} \affiliation{Sternberg Astronomical Institute, M.V. Lomonosov Moscow State University, 13, Universitetskij pr., 119234, Moscow, Russia}

\author[0000-0002-7424-9891]{Justin M. Wittrock} \affiliation{Department of Physics and Astronomy, George Mason University, 4400 University Drive, Fairfax, VA 22030, USA}

\author[0000-0003-1963-9616]{Douglas~A.~Caldwell} \affiliation{SETI Institute, Mountain View, CA 94043, USA}

\author[0000-0002-2482-0180]{Zahra~Essack} \affiliation{Department of Earth, Atmospheric and Planetary Sciences, MIT, Cambridge, MA 02139, USA} \affiliation{Department of Physics and Kavli Institute for Astrophysics and Space Research, MIT, Cambridge, MA 02139, USA}

\author[0000-0002-4715-9460]{Jon~M.~Jenkins} \affiliation{NASA Ames Research Center, Moffett Field, CA 94035, USA}

\author[0000-0003-1309-2904]{Elisa V. Quintana} \affiliation{NASA Goddard Space Flight Center, Greenbelt, MD 20771, USA}

\author[0000-0003-2058-6662]{George R. Ricker} \affiliation{Department of Physics and Kavli Institute for Astrophysics and Space Research, MIT, Cambridge, MA 02139, USA}

\author[0000-0001-6763-6562]{Roland~Vanderspek} \affiliation{Department of Physics and Kavli Institute for Astrophysics and Space Research, MIT, Cambridge, MA 02139, USA}

\author[0000-0002-6892-6948]{S.~Seager}\affiliation{Department of Physics and Kavli Institute for Astrophysics and Space Research, MIT, Cambridge, MA 02139, USA} \affiliation{Department of Earth, Atmospheric and Planetary Sciences, MIT, Cambridge, MA 02139, USA} \affiliation{Department of Aeronautics and Astronautics, MIT, 77 Massachusetts Avenue, Cambridge, MA 02139, USA}

\author[0000-0002-4265-047X]{Joshua N.\ Winn}\affiliation{Department of Astrophysical Sciences, Princeton University, Princeton, NJ 08544, USA}
\email{eturtelboom@berkeley.edu}

\begin{abstract}
Multi-planet systems are valuable arenas for investigating exoplanet architectures and comparing planetary siblings. TOI-1246 is one such system, with a moderately bright K dwarf ($\rm{V=11.6,~K=9.9}$) and four transiting sub-Neptunes identified by \tess{} with orbital periods of  $4.31~\rm{d},~5.90~\rm{d},~18.66~\rm{d}$, and $~37.92~\rm{d}$. We collected \totalRVs{} radial velocity observations with Keck/HIRES and TNG/HARPS-N to measure planet masses. We refit the 14 sectors of TESS photometry to refine planet radii ($\rm{2.97 \pm 0.06~R_\oplus},\rm{2.47 \pm 0.08~R_\oplus}, \rm{3.46 \pm 0.09~R_\oplus}$, $\rm{3.72 \pm 0.16~R_\oplus}$), and confirm the four planets. We find that TOI-1246 e is substantially more massive than the three inner planets (\massc $M_\oplus$, \massd $M_\oplus$, \massb $M_\oplus$, \masse $M_\oplus$). The two outer planets, TOI-1246 d and TOI-1246 e, lie near to the 2:1 resonance ($\rm{P_{e}/P_{d}=2.03}$) and exhibit transit timing variations. \mbox{TOI-1246} is one of the brightest four-planet systems, making it amenable for continued observations. It is one of only six systems with measured masses and radii for all four transiting planets. The planet densities range from $\rm{0.70 \pm 0.24}$ to  $3.21 \pm 0.44 \rm{g/cm^3}$, implying a range of bulk and atmospheric compositions. We also report a fifth planet candidate found in the RV data with a minimum mass of 25.6 $\pm$ 3.6 $\rm{M_\oplus}$. This planet candidate is exterior to TOI-1246 e with a candidate period of 93.8 d, and we discuss the implications if it is confirmed to be planetary in nature. 
\end{abstract}


\section{Introduction} \label{sec:intro}

While we have known for centuries that our own Solar System hosts multiple planets, it was only in the late twentieth century that we confirmed the existence of multiple planets around other stars. In fact, the first exoplanets ever discovered were two planets orbiting the pulsar PSR 1257+12 \citep{wolszczan+frail1992}, i.e.\ the first known exoplanetary system was also the first known multi-planet system. In the years after this finding, ground-based radial velocity (RV) surveys found additional planets orbiting known planet hosts \citep{fischer+2002,butler+1999}, and methods were developed to appropriately model multi-planet RV signals \citep{wright+howard2009}. Still later, the prevalence of multi-planet systems was brought to light by large surveys of transiting planets such as the \kep{} and \tess{} missions. Of the 2385 confirmed planets discovered using the transit method by the \kep{} mission, 1156 were found to be in 458 multi-planet systems\footnote{NASA Exoplanet Archive \citep{NEA}, \url{exoplanetarchive.ipac.caltech.edu}, accessed on 8 November 2021}. The \tess{} mission is continuing to expand our understanding of exoplanets by searching for bright, nearby stars that host transiting planets. About $\rm{15\%}$ of \tess{} Objects of Interest (TOIs) are predicted to be in systems with multiple transiting planets \citep{huang+2018}. While 34 \tess{} systems host multiple confirmed planets, only four systems host four or more transiting planets (HD 108236: \citealt{bonfanti+2020}, L98-59: \citealt{demangeon+2021}. TOI 178: \citealt{leleu+2021}, and TOI 561: \citealt{lacedelli+2021, weiss+2021}) . In this work we add to this growing, but currently small, sample. 

This paper is the eleventh (TKS-XI) in a series of papers by the \tess-Keck Survey (TKS). TKS is a collaboration spanning several institutions that pools time on the Keck-I telescope on Maunakea. TKS conducts spectroscopy using the High Resolution Echelle Spectrograph (HIRES, \citealt{vogt+1994}) of TOIs in order to determine planet masses and host star properties for a variety of science goals. TKS-0 \citep{tks0} is a more comprehensive description of the TKS program science goals and target selection. 

The TOI-1246 system hosts four transiting sub-Neptune-sized planets orbiting a K dwarf: TOI-1246 b (formerly TOI-1246.02), TOI-1246 c (formerly TOI-1246.03), TOI-1246 d (formerly TOI-1246.01), and TOI-1246 e (formerly TOI-1246.04). The four transiting planets lie within 0.5 AU of the host star ($P_{b} = 4.31~\rm{d}$, $P_{c} = 5.90~\rm{d}$, $P_{d} = 18.66~\rm{d}$, $P_{e} = 37.92~\rm{d}$). The inner three planets are smaller than the outermost transiting planet ($\rm{R_{b} =~2.97 \pm 0.06~R_\oplus}$, $\rm{R_{c} =~2.47 \pm 0.08~R_\oplus}$, $\rm{R_{d} =~3.46 \pm 0.09~R_\oplus}$, $\rm{R_{e} =~3.72 \pm 0.16~R_\oplus}$). By studying multiple planets orbiting a single host star, we can study planets with a shared formation and evolutionary history.

TOI-1246 was observed as part of the TKS sub-program to obtain follow-up observations and determine masses for planets in multi-planet systems observed by \tess{}. This paper also includes RV data collected using TNG/HARPS-N. The paper is structured as follows. In \S \ref{sec:obs} we describe the \tess{} photometry, follow-up spectroscopic, photometric, and imaging observations used in our analysis of the TOI-1246 system. In \S \ref{sec:star} we characterize the host star and investigate stellar rotation and activity. In \S \ref{sec:analysis}, we refit the \textit{TESS} photometry to refine transit parameters, and analyze the two RV data sets in order to determine planet masses and constrain planetary orbits. We also measure transit timing variations (TTVs) and investigate system stability. Finally, in \S \ref{sec:results+discussion} we discuss the planet masses and densities in the context of other multi-planet systems, as well as the suitability of this system for further studies involving atmospheric characterization. We also discuss the system architecture and the possible explanations and implications of the fifth planet candidate in \S \ref{sec:results+discussion}, and we conclude this work in \S \ref{sec:conclusion}.

\section{Observations} \label{sec:obs}
In this section we describe the three types of observations of TOI-1246: photometric data (from the \textit{TESS} mission and seeing-limited photometry collected by the \tess{} SG1 Working Group), spectroscopic data (reconnaissance and precision spectra), and high-resolution imaging data (speckle, lucky, and AO). Additional figures describing the observations discussed here can be found on ExoFOP-TESS\footnote{\citet{exofop}, \url{https://exofop.ipac.caltech.edu/tess/target.php?toi=1246}}.
\subsection{Photometric Data} 
\subsubsection{\tess{} Observations}
\begin{figure}
    \centering
    \includegraphics[width=\columnwidth]{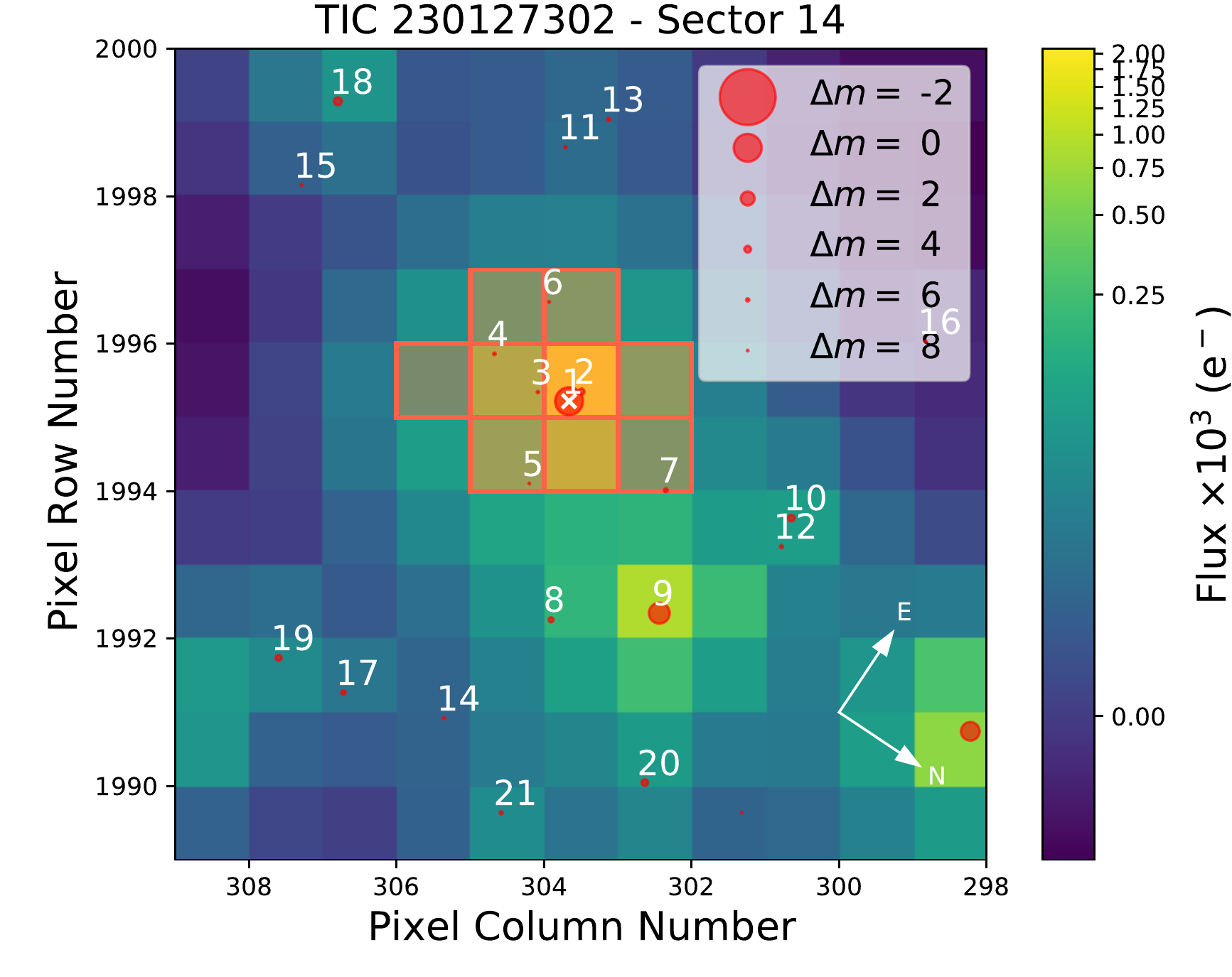}
    \caption{Sector 14 \tess{} Target Pixel File for TOI-1246 with the SPOC pipeline aperture overlaid in red. TOI-1246 is indicated with a white cross and labelled as target 1. The \textit{Gaia} DR2 sources in the field are shown by red dots, scaled in size to the difference in magnitude with TOI-1246. The \tess{} pixel scale is 21 arcsec $\rm{pixel^{-1}}$.}
    \label{fig:tpf}
\end{figure}

TOI-1246, also known as TIC 230127302 in the \tess{} Input Catalog (TIC, \citealt{tic7_catalog}), was selected for observation at 2-minute cadence for a total of 12 sectors in Cycle 2 of the \tess{} mission (sectors 14 - 17 and 19 - 26, 18 July 2019 - 2 November 2019 and 27 November 2019 - 4 July 2020). The target lies in \tess{}'s Northern Continuous Viewing Zone, and was observed for a total of 327 days using either camera 3 or 4. Figure \ref{fig:tpf} shows the Target Pixel File  for TOI-1246 with the \tess{} aperture and nearby stars identified by \textit{Gaia} shown for context. Through additional time series observations (see Section \ref{sec:timeseries}), we confirm the transit events are on-target, and imaging observations indicate that these stars do not significantly contaminate the flux observed (see Section \ref{sec:imaging}).

The system is being re-observed for a further 273 days in \tess{} Cycle 4 in sectors 40-41, 47, and 49-55 (24 June 2021 - 8 August 2021, 30 December 2021 - 28 January 2022, 26 February 2022 - 1 September 2022). TOI-1246 has also been selected for observation at 20-second cadence in Sectors 40 and 41 through the Guest Investigator Programs\footnote{\url{https://heasarc.gsfc.nasa.gov/docs/tess/approved-programs.html\#cycle-4}} G04039 (PI: Davenport) and G04242 (PI: Mayo). These continued observations will extend the photometric baseline to a total of 1141 days. We make use of \tess{} photometric data up to and including sector 41 in this work, and will analyze the additional \tess{} data in a future work.

The four transiting planets in this system were initially detected by the SPOC pipeline \citep{spoc} and released as TOIs on 17 October 2019 (for TOI-1246 b and d) and 15 November 2019 (TOI-1246 c). TOI-1246 e ($\rm{P = 37.92 d}$) was labelled as a Community TOI known as CTOI 230127302 e by Martti Holst Kristiansen\footnote{\citet{exofop}, \url{https://exofop.ipac.caltech.edu/tess/view_tag.php?tag=18072}} (Brorfelde Observatory, Denmark) before being promoted to a TOI on 16 July 2020. 

\begin{figure*}
    \centering
    \includegraphics[width=\textwidth]{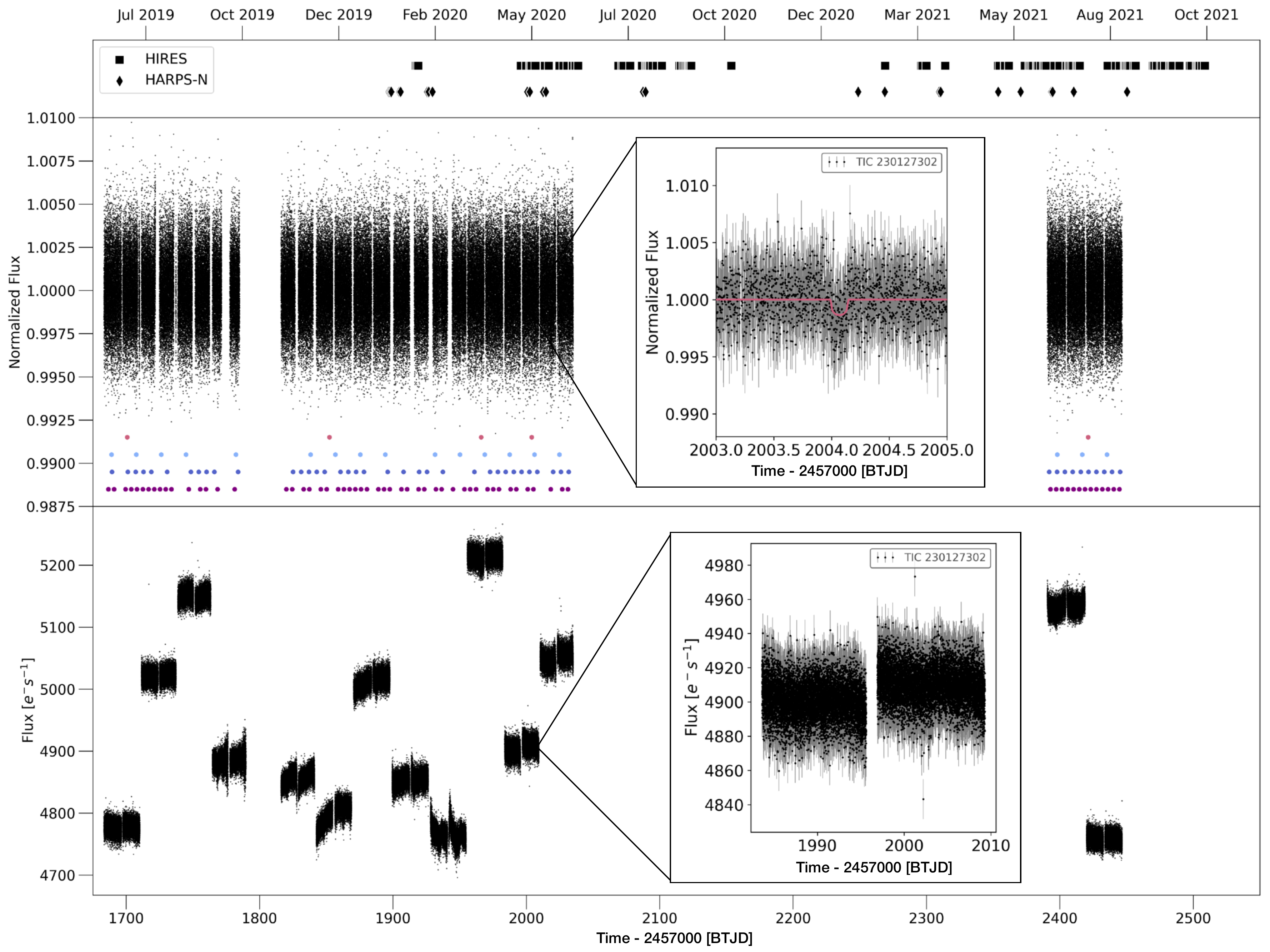}
    \caption{\textit{Top panel:} The times of Keck/HIRES and TNG/HARPS-N RV observations of TOI-1246, which were contemporaneous with \tess{} observations. Times are shown in BTJD which is defined as $\rm{BJD_{TDB}}$ - 2457000. $\rm{BJD_{TDB}}$ is the Barycentric Julian Date in the Barycentric Dynamical Time standard. \textit{Middle panel:} Flattened and normalized light curve of TOI-1246 derived from \tess{} PDCSAP photometry using the Lightkurve package \citep{lightkurve}. The times of full transits are indicated with dots below the light curve in various colours (TOI-1246 b: purple, TOI-1246 c: blue, TOI-1246 d: light blue, TOI-1246 e: pink). A zoomed-in inset is shown of a single transit of TOI-1246 e with the best-fit model overlaid (see Section \ref{sec:phot}). \textit{Lower panel:} Raw \tess{} SAP photometry of TOI-1246 with a zoomed-in inset of sector 23 to illustrate intra-sector variability.}
    \label{fig:lc_transits}
\end{figure*}

Figure \ref{fig:lc_transits} shows the raw and the normalized and flattened (using a Savitzky-Golay filter, \citealt{savtizky+golay1964}) photometric light curves for TOI-1246. As of Sector 41, TESS has observed 64 transits of TOI-1246 b, 46 transits of TOI-1246 c, 18 transits of TOI-1246 d, and 6 transits of TOI-1246 e; the planet ephemerides suggest that additional transits occurred during gaps in the TESS photometry. We used the PDCSAP\_FLUX \citep{stumpe+2012, stumpe+2014, smith+2012} times series observations in our analysis of this system.

\subsubsection{Time Series Observations} \label{sec:timeseries}
We conducted ground-based photometric follow-up observations of TOI-1246 as part of the \tess{} Follow-up Observing Program\footnote{\url{https://tess.mit.edu/followup}} \citep[TFOP;][]{collins:2019} to confirm the transit-like signals on-target or identify near-by eclipsing binaries (NEBs) as potential sources of the \textit{TESS} detection. We used the {\tt TESS Transit Finder}, which is a customized version of the {\tt Tapir} software package \citep{Jensen:2013}, to schedule our transit observations. All photometric data were extracted using {\tt AstroImageJ} \citep{Collins:2017}, except the Dragonfly observations as described below.

We observed a transit egress of TOI-1246 d using the MORAVIAN-G4 9000	camera and the \textit{Ic} filter on the 0.4 m telescope at the Observatori Astronòmic Albanyà (Girona, Spain). We took these observations on 17 May 2020 UT, and reported a tentative detection of an event that was on-target and in line with the predicted transit depth (measured as $1429 \pm 67$ ppm by \tess{}). Egress occurred approximately 37 min later than predicted (see Section \ref{sec:ttvs} for a discussion of TTVs), and we cleared several nearby stars of being NEBs. 

We observed a transit of TOI-1246 b using the SBIG STX-16803 + FW-7 camera and the R filter at the George Mason University Observatory on 29 October 2019 between 00:59:34 and 04:50:55 UT. Due to challenging weather conditions and the target's shallow transit depth (measured as $1104 \pm 46$ ppm by \tess{}), this observation resulted in a non-detection. However, we ruled out the possibility that one of the near-by stars is an NEB. We observed the target for a total 231.4 minutes.

We also observed a transit of TOI-1246 b using ASA 1 m f/6 Ritchey-Chrétien telescope, Spectral Instruments 1110S camera and ACE filter wheel/guide camera with the \textit{Ic} filter at the US Air Force Academy on 6 March 2020 UT. We conclusively detected the transit event with a depth of 0.897 ppt. We note that the observed ingress commenced about 3 minutes later than predicted, and egress ended about 5 minutes early. 

We observed another transit of TOI-1246 b using the Las Cumbres Observatory Global Telescope \citep[LCOGT;][]{Brown:2013} 1 m Sinistro instrument at McDonald Observatory on 14 May 2020 UT. We used the \textit{zs} filter and observed the target for a total of 270 minutes, covering the full transit. The images were calibrated by the standard LCOGT {\tt BANZAI} pipeline \citep{McCully:2018}. These observations confirmed that the transit signal was on-time and on-target.

We observed a fourth transit of TOI-1246 b on 28 May 2021 UT, using the Dragonfly Telephoto Array (DRA). The DRA, housed at the New Mexico Skies telescope hosting facility, is a remote telescope consisting of an array of small telephoto lenses roughly equivalent to a 1.0 m refractor \citep{danieli+2020}. DRA has a SBIG STF-8300M detector with a 156’ x 114' field of view. Simultaneous observations of TOI-1246 were conducted in g' and r' bands with 44 s exposure times. We observed the target for 134 minutes pre-transit, 131 minutes in-transit, and for 97 minutes post-transit. The data were reduced and analyzed with a custom differential aperture photometry pipeline designed for multi-image processing and analysis. We observed the full transit and made a marginal detection of an on-time and on-target transit, and report two contaminating sources which are fainter by 3.81 and 6.65 magnitudes, respectively.

We observed a transit of TOI-1246 c using the FLI 16803 camera  and \textit{Ic} filter at Adams Observatory on 14 July 2020 UT. We used an exposure time of 180 s and observed the target for 241 minutes. We were unable to detect the transit signal on target due to the shallow transit depth, but cleared two nearby stars of being NEBs. 

\subsection{Imaging Observations} \label{sec:imaging}
\subsubsection{Speckle and Lucky Observations}
Stars with small projected separations from an exoplanet host star can create a false-positive transit signal. In addition, ``third-light” flux from the close companion star can lead to an underestimated planetary radius, an incorrect mean density, and imprecise stellar parameters (see \citealt{ciardi+2015, furlan+2017}). The glare from a close companion can also cause a non-detection of the transits of small planets residing within the same system \citep{lester+2021}. Thus, to search for close-in companions unresolved in TESS or other ground-based follow-up observations, we obtained high-resolution imaging observations of TOI-1246.

TOI-1246 was observed on 25 June 2021 UT using the ‘Alopeke speckle instrument on the Gemini North 8 m telescope \citep{scott+2021}. ‘Alopeke provides simultaneous speckle imaging in two bands (562 nm and 832 nm) and produced a reconstructed image with robust contrast limits on companion detections (e.g., \citealt{howell+2016}). Seven sets of 1000 $\times$ 0.06 s exposures were collected and subjected to Fourier analysis in our standard reduction pipeline \citep[see][]{howell+2011}. The Fourier transform of the summed autocorrelation of each set of images is used to make a fringe image of the target, which is then used to reconstruct the image. We find that TOI-1246 is a single star with no companion fainter than the target star by 5-7 magnitudes from 0.1\arcsec-1.2\arcsec (i.e. 17-203 AU). 

We also observed TOI-1246 in the \textit{Ic} band on 24 January 2021 UT with the SPeckle Polarimeter (SPP; Safonov et al. 2017) on the 2.5 m telescope at the Caucasian Observatory of Sternberg Astronomical Institute (SAI) of Lomonosov Moscow State University. SPP uses an Electron Multiplying CCD Andor iXon 897 as a detector, and we used the atmospheric dispersion compensation. The detector has a pixel scale of 20.6 mas/pixel, the angular resolution is 89 mas, and the field of view is 5\arcsec x5\arcsec centered on the star. The power spectrum was estimated from 4000 frames with 30 ms exposures. We did not detect any stellar companions brighter than $\Delta$mag = 3.8 and 5.5 at 0.2\arcsec and 0.5\arcsec, respectively.

Finally, we observed TOI-1246 with the AstraLux instrument (\citealt{hormuth08}) installed at the 2.2 m telescope in the Calar Alto Observatory (Almer\'{i}a, Spain) under average weather and atmospheric conditions (seeing around 1\arcsec) on the night of 25 February 2020 UT with the SDSSz filter. AstraLux uses the lucky imaging technique to obtain thousands of short exposure frames and selects a few percent of these frames with the best Strehl ratio \citep{strehl1902}. This process is entirely done by the instrument pipeline. We obtained 95\,400 frames with an exposure time of 20 ms each and selected the best 10\% for a final effective exposure time of 190.8\,s. We used the finally stacked image to obtain the contrast curve by using the \texttt{astrasens} code \citep{lillobox+2012,lillo-box14b}. The result provides a contrast of $\Delta z=5$~mag for separations above 0.3\arcsec and a maximum contrast of $\Delta z=3$~mag for 0.1\arcsec. We found no additional sources in the field of view of the instrument ($3\arcsec\times 3\arcsec$ in this setup) within these sensitivity limits.

\subsubsection{Adaptive Optics Observations} \label{sec:ao}
\begin{figure}
    \includegraphics[width=\columnwidth]{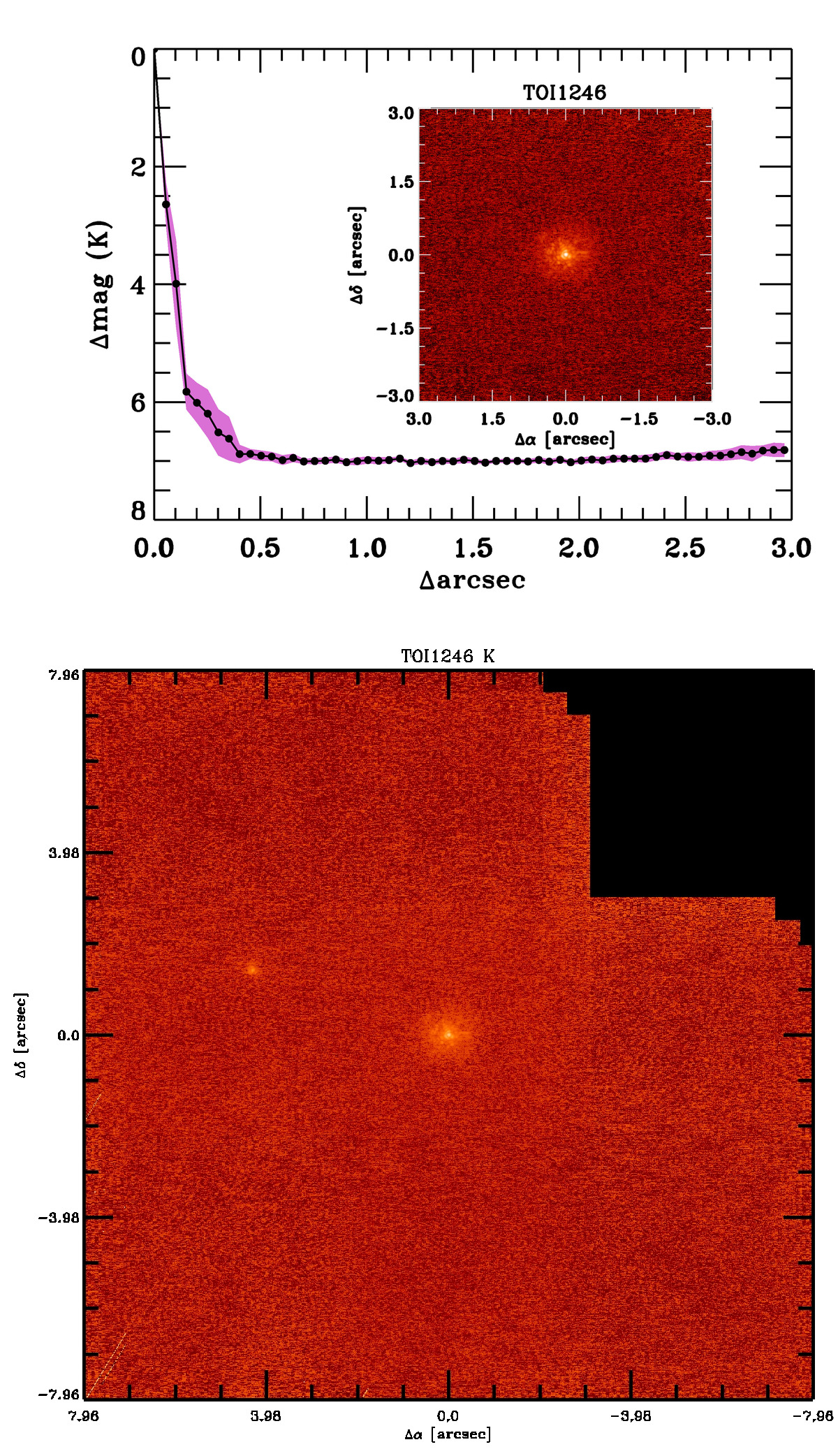}
    \caption{\textit{Top:} Sensitivity curve and zoomed-in image in K band of TOI-1246 using Keck/NIRC2. \textit{Bottom:} AO image of TOI-1246 using Keck/NIRC2. We see a companion at $\sim4\arcsec$ to the NE of TOI-1246, which does not significantly affect our analysis.}
    \label{fig:ao}
\end{figure}

We observed TOI-1246 with the NIRC2 imager \citep{wizinowich+2000} on the Keck-II telescope on 28 September 2020 UT. We took observations in the narrow camera mode (0 d\arcsec/pixel) with a 1024 x 1024 pixel FOV, and used a three-point dither pattern to avoid the noisy fourth quadrant of the detector. We used the $K$ filter for a total integration time of 4.5 s and the $J$ filter for a total integration time of 9 s. All of the data were processed and analyzed with a custom set of IDL tools, and the science frames were flat-fielded and sky-subtracted. The sensitivity curve and image resulting from these observations are shown in Figure \ref{fig:ao}. 

We note that there is a proper motion companion star (TIC 230127303) detected by \textit{Gaia} that is 4\arcsec to the NE of TOI-1246. Based on the measured distance of TOI-1246 and this companion ($169$ pc), these two stars are separated by $\sim750$ AU. This separation is well within the 21\arcsec  $\times$ 21\arcsec \tess{} pixels, and so the flux from this companion star affects the photometric light curve for TOI-1246. The companion is 3.8 magnitudes fainter in the \tess{} bandpass (magnitudes sourced from the TIC \citealt{tic7_catalog}), and implies a radius correction factor of 1.015 following \citet{furlan+2017}. This is smaller than the errors associated with our derived radii (see Section \ref{sec:phot}), and so does not affect our conclusions. Furthermore, this stellar companion does not cause a detectable trend in the RV data due to its large separation from TOI-1246 (see Section \ref{sec:RV} for details). 

\subsection{Spectroscopic Data}

\subsubsection{Tillinghast/TRES Spectra}
We observed two reconnaissance spectra of TOI-1246 on 20 October 2019 UT and on 20 February 2020 UT with the Tillinghast Reflector Echelle Spectrograph (TRES; \citealt{tres}) located at the Fred Lawrence Whipple Observatory (FLWO) in Arizona, USA. TRES is a fiber-fed spectrograph with a resolving power of 44,000. The spectra were extracted as described in \citet{buchhave+2010} and were then cross-correlated, order by order, against each other using the strongest observation as a template. This multi-order analysis revealed a velocity variation of 27 m/s between the two spectra, similar to the uncertainty of the measurement. We found no significant velocity variation between the two observations, indicating the star is well-suited for precise RV observations.
We also derived stellar parameters using the Stellar Parameter Classification (SPC; \citealt{buchhave+2012}) tool. SPC cross correlates an observed spectrum against a grid of synthetic spectra based on Kurucz atmospheric models \citep{kurucz1992}. The weighted average results derived with $\rm{T_{eff}}$, log(g), [M/H], and $\rm{vsin\textit{i}}$ as free parameters are reported in Table \ref{tab:stellar}.

\subsubsection{Keck/HIRES Spectra}
TKS obtained high-resolution spectra of TOI-1246 using the HIRES spectrograph on the Keck-I telescope on Maunakea. We collected \HIRESRVs{} spectra between November 2019 and October 2021. HIRES operates between 360 and 800 nm, and TOI-1246 was observed using the red cross-disperser, C2 decker (14\arcsec  $\times$ 0.861\arcsec, R = 60,000), and with a median exposure time of 1004 s. The \HIRESRVs{} RV observations were taken with a warm ($50\degree C$) iodine cell in the light path for wavelength calibrations as per \citet{butler+1996}. Two further higher-resolution spectra were taken without the iodine cell in the light path (`iodine-out') in November 2019 and June 2020 in order to obtain a spectral template, using the B3 decker (14''  $\times$ 0.574'', R = 72,000). The spectra were reduced using the standard procedures described in \citet{howard+2010}. The RVs, RV errors, and Mount Wilson S-Index (a proxy for stellar activity derived from Ca II H \& K lines, e.g. \citealt{isaacson+fischer2010}) values collected on Keck/HIRES are reported in Table \ref{tab:rvdata}.

\subsubsection{TNG/HARPS-N Spectra}
A further \HARPSRVs{} high-resolution spectra of TOI-1246 were collected using the HARPS-N instrument installed at the Telescopio Nazionale Galileo (TNG) of Roque de los Muchachos Observatory in La Palma, Spain\footnote{13 spectra were obtained from the Spanish CAT19A\textunderscore162 program (PI: Nowak), 7 spectra from ITP19\textunderscore1 program (PI: Pall\'e) and 9 spectra from CAT21A\textunderscore119 program (PI: Nowak).}. These observations were taken between February 2020 and September 2021. The exposure time was set to \mbox{1800 - 3600 s}, based on weather conditions and scheduling constraints, leading to a SNR per pixel of 25--56 at 5500~\AA. We used the {\tt serval} code \citep{zechmeister+2018} to measure relative RVs by template-matching, and also to derive the chromatic index (CRX), differential line width (dLW), H$\alpha$, and sodium Na~D1 $\&$ Na~D2 indexes. Doppler measurements and spectral activity indicators (CCF\_FWHM, CCF\_CTR, BIS and Mount Wilson S-index) were measured using an online version of the DRS, the YABI tool\footnote{Available at \url{http://ia2-harps.oats.inaf.it:8000}.}, by cross-correlating the extracted spectra with a K5 mask \citep{baranne+1996}. The RVs, RV errors, and values of the Mount Wilson S-Index collected on TNG/HARPS-N are reported in Table \ref{tab:rvdata}.

\section{Stellar Parameters} \label{sec:star}
We used several methods to characterize TOI-1246. We report these values in Table \ref{tab:stellar}, and indicate with an asterisk the preferred values that we used where multiple values were obtained for a given parameter (we note the multiple values are consistent). We used the \texttt{SpecMatch-Synthetic}\footnote{\url{ github.com/petigura/specmatch-syn}} code \citep{specmatch-synth} to fit sections of the iodine-out optical spectrum collected with Keck/HIRES using forward modeling by interpolating between a grid of model spectra from \citet{coelho+2005}. We use this method to derive the effective temperature ($\rm{T_{eff}}$ = 5151 $\pm$ 100 K), stellar radius (\mbox{$\rm{R_*}$ = 0.86 $\pm$ 0.05 $\rm{R_\odot}$}), surface gravity (\mbox{log(g) = 4.4 $\pm$ 0.1}), and metallicity (\mbox{[Fe/H] = 0.17 $\pm$ 0.06}) of TOI-1246. 

\begin{figure}
    \centering
    \includegraphics[width=\columnwidth]{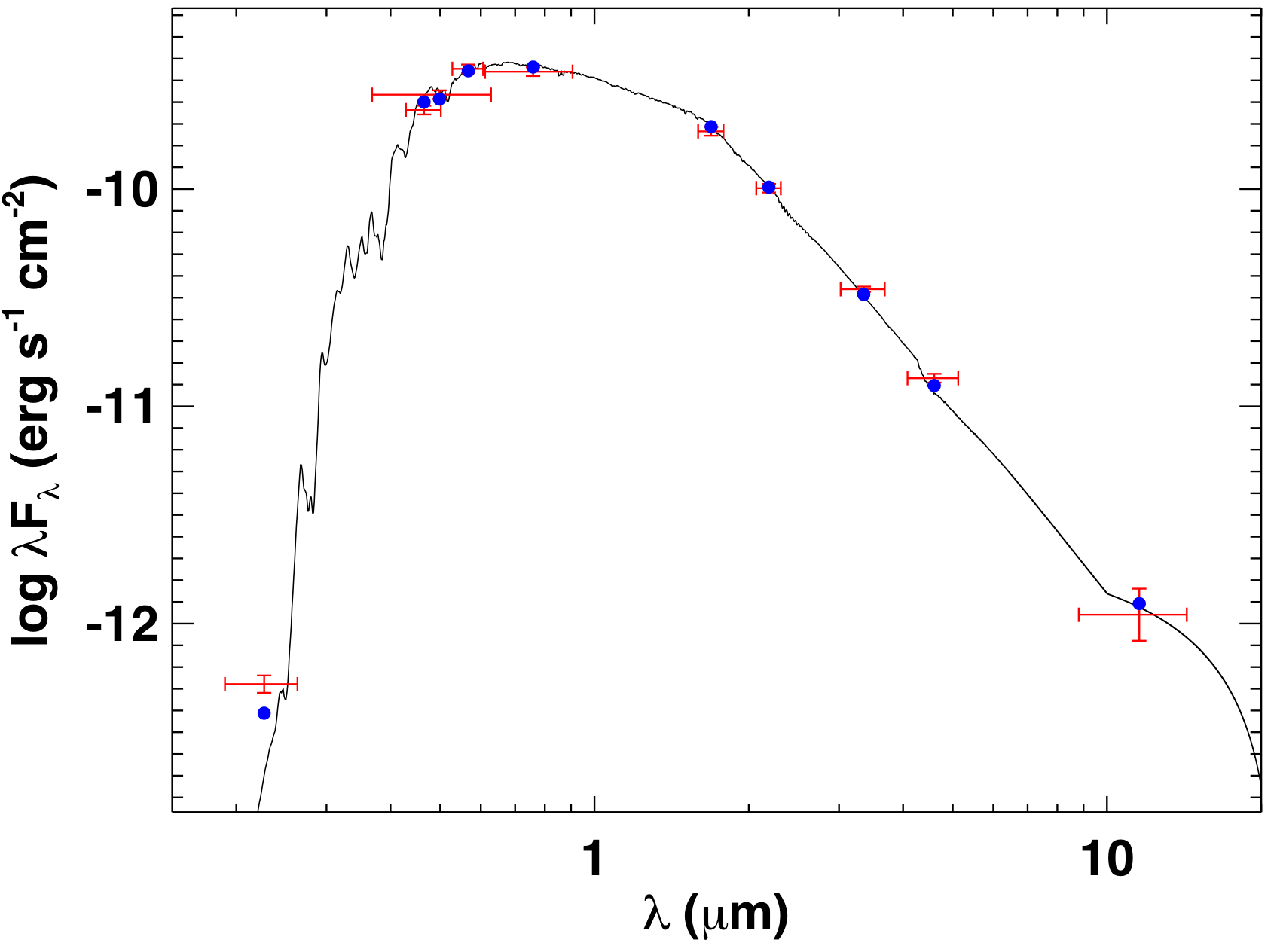}
    \caption{Spectral energy distribution of TOI-1246. Red symbols represent the observed photometric measurements, and the horizontal bars represent the effective width of the passband. Blue symbols are the model fluxes from the best-fit Kurucz atmosphere model (black). We observe some NUV excess that implies moderate stellar activity for this star. \label{fig:sed}}
\end{figure}

We also analyzed the broadband spectral energy distribution (SED) of the star together with the {\it Gaia\/} EDR3 parallax \citep[with no systematic offset applied; see, e.g.,][]{StassunTorres:2021}, in order to measure the stellar radius, following the procedures described in \citet{Stassun:2016,Stassun:2017,Stassun:2018}. We obtained the $B,V$ magnitudes from {\it APASS}, the $J,H,K_S$ magnitudes from {\it 2MASS}, the W1--W4 magnitudes from {\it WISE}, the $G_{\rm BP},G_{\rm RP}$ magnitudes from {\it Gaia}, and the NUV magnitude from {\it GALEX}. The available photometry spans the full stellar SED from 0.2 to 22~$\mu$m (see Figure~\ref{fig:sed}). 

We performed a fit using Kurucz stellar atmospheric models \citep{kurucz1993}, with the effective temperature ($T_{\rm eff}$), metallicity ([Fe/H]), and surface gravity ($\log(g)$) adopted from the spectroscopic analysis. The remaining free parameter is the extinction $A_V$, which we limited to the maximum line-of-sight value from the Galactic dust maps of \citet{Schlegel:1998}. The resulting fit has a best-fit $A_V = 0.11 \pm 0.02$ and a reduced $\chi^2$ of 2.7, but we note that the {\it GALEX} NUV flux indicates a moderate level of activity. Integrating the (unreddened) model SED gave the bolometric flux at Earth, $F_{\rm bol} = 5.56 \pm 0.20 \times 10^{-10}$ erg~s$^{-1}$~cm$^{-2}$. Taking the $F_{\rm bol}$ and $T_{\rm eff}$ together with the {\it Gaia\/} parallax, we derived a stellar radius, $R_\star = 0.895 \pm 0 c8$~R$_\odot$. In addition, we estimated the stellar mass from the empirical relations of \citet{Torres:2010}, giving $M_\star = 0.93 \pm 0.05$~M$_\odot$, which is consistent with the estimate of $M_\star = 0.74 \pm 0.18$~M$_\odot$ obtained directly from $R_\star$ and the spectroscopic $\log(g)$. 

We used the stellar NUV excess (see Figure~\ref{fig:sed}) to estimate a rotation period and age via empirical rotation-activity-age relations. The observed NUV excess implies a chromospheric activity of $\log R'_{\rm{HK}} = -4.98 \pm 0.05$ following \citet{Findeisen:2011}, which is consistent with the $\log R'_{\rm{HK}}$  measured with Keck/HIRES ($\log R'_{\rm{HK}} = -5.10 \pm 0.15$). We also measured the Mount Wilson S-Index to be $0.150 \pm 0.001$ using Keck/HIRES, which indicates low chromospheric activity \citep{wilson_sindex}. The $\log R'_{\rm{HK}}$ value implies a stellar rotation period of $P_{\rm rot} = 47 \pm 3$~d according to the empirical relations of \citet{Mamajek:2008}. However, the values of $\log R'_{\rm{HK}}$ and B-V (0.943) found for TOI-1246 are at the edge of (or beyond) the range considered in \citet{Mamajek:2008} ($\log R'_{\rm{HK}}$ < -5, and $\rm{0.5 < B-V < 0.9}$), and so the uncertainty is likely underestimated. The NUV estimated activity also implies an age of $6.2 \pm 0.8$~Gyr via the empirical relations of \citet{Mamajek:2008}, but this error is also likely underestimated.

We also attempted to obtain a rotation period measurement directly from the data. We first performed a un-informed search for stellar rotation in the \tess{} 2- and 30- minute cadence light curves. We could find no convincing rotation period using a Lomb-Scargle periodogram \citep{lomb1976, scargle1982}, autocorrelation function \citep{mcquillan+2013}, or wavelet decomposition \citep{torrence+compo1998, mathur+2020}. However, rotation periods longer than 13.7 days have eluded detection in \tess{} light curves due to the \mbox{27 day} sector length and thermal effects on the detector sensitivity related to the data down-link at the midpoint of each sector \citep[e.g.][]{cantomartins+2020, avallone+2021}. Therefore, we passed the light curve's wavelet decomposition through the convolutional neural network of \citet{claytor+2021}. The neural network is trained using wavelet decomposition of simulated rotational light curves with real \tess{} systematics and noise. Given an input image of the wavelet transform, the neural network predicts a rotation period and a heuristic (but not statistical) uncertainty. Based on ensemble recovery, if the uncertainty is less than about 35\% of the predicted period, the period is likely to be real. We found no reliable rotation period with this method.

In the absence of photometric rotation signatures, we put theoretical constraints on the stellar rotation period using the stellar model-fitting tools in \texttt{kiauhoku} \citep{claytor+2020}. We used a MCMC routine to fit the models of \citet{vansaders+2013} and \citet{vansaders+2016} to the star's effective temperature, metallicity, and surface gravity, yielding rotation period predictions following three different braking laws. The different braking laws vary combinations of starting condition and stalled braking behavior at a critical Rossby number \citep{vansaders+2016}. Regardless of the braking law used, we predicted a rotation period of \mbox{$38^{+7}_{-14}$ d} for TOI-1246, which is consistent with the prediction from \citet{Mamajek:2008}. Both results are consistent with the rotation period estimate from the spectroscopic  $\rm{vsin(i)}$ and $\rm{R_*}$ of 42 $\pm$ 40 days, due to the large error on $\rm{vsin(i)}$. Overall, TOI-1246 is a relatively old, low-activity and slightly metal rich K dwarf. 

\begin{deluxetable*}{lcccc}[h]
\tablecaption{TOI-1246 Stellar Parameters \label{tab:stellar}}
\tablehead{\colhead{Parameter} & \colhead{Value} & \colhead{Error} & \colhead{Source} & \colhead{Adopted?}} 
\startdata
Other Names & TIC 230127302, & & TIC \citet{tic7_catalog} & - \\
& TYC 4423-02107-1 & & TYCHO \citet{tycho} & - \\
& Gaia DR2 1650110904522335744 & & \textit{Gaia} DR2 \citet{gaiadr2} & - \\
Right Ascension (hh:mm:ss) & 16:44:27.81 & &  TIC v8.2 & - \\
Declination (hh:mm:ss) & +70:25:47.97 & & TIC v8.2 & - \\
V magnitude & 11.632 & 0.024 & TIC v8.2 & - \\
TESS magnitude & 11.1802 & 0.0061 & TIC v8.2 & - \\
J magnitude & 10.294 & & TIC v8.2 & - \\
K magnitude & 9.907 & 0.036 & TIC v8.2 & - \\
\textit{Gaia} magnitude & 11.7248 & 0.0002 & \textit{Gaia} DR2 \citep{gaiadr2} & - \\
Parallax (mas) & 5.847 & 0.011 & \textit{Gaia} DR2 \citep{gaiadr2} & - \\
RA proper motion (mas/yr) & -48.024 & 0.047 & \textit{Gaia} DR2 \citep{gaiadr2} & - \\
Dec proper motion (mas/yr) & 81.928 & 0.051 & \textit{Gaia} DR2 \citep{gaiadr2} & - \\
\hline
Radius $\rm{(R_\odot)}$ & 0.86 & 0.05 &  SpecMatch-Synthetic \citep{specmatch-synth} &  Y \\
Radius $\rm{(R_\odot)}$ & 0.895 & 0.038 &  Calculated using $\rm{F_{bol}}$, $\rm{T_{eff}}$, and parallax & - \\
Radius $\rm{(R_\odot)}$ & 0.876 & 0.051 &  TIC v8.2 & - \\
\hline
Mass $\rm{(M_\odot)}$ & 0.87 & 0.03 &  SpecMatch-Synthetic \citep{specmatch-synth} & Y \\
Mass $\rm{(M_\odot)}$ & 0.93 & 0.05 & Calculated using \citet{Torres:2010} & - \\
Mass $\rm{(M_\odot)}$ & 0.74 & 0.18 &  Calculated using $\rm{log(g) and R_*}$ & - \\
Mass $\rm{(M_\odot)}$ & 0.868 & 0.105 &  TIC v8.2 & - \\
Mass $\rm{(M_\odot)}$ & 1.12 & 0.16 & Photometric Fit (Section \ref{sec:phot}) & - \\
\hline
$\rm{T_{eff} (K)}$ & 5141 & 122 &  TIC v8.2 & - \\
$\rm{T_{eff} (K)}$ & 5151 & 100 &  SpecMatch-Synthetic \citep{specmatch-synth} & Y \\
$\rm{T_{eff} (K)}$ & 5217 & 50 &  SPC \citep{buchhave+2012}  & - \\
\hline
$\rm{log(g)}$ & 4.4 & 0.1 & SpecMatch-Synthetic \citep{specmatch-synth} & Y \\
$\rm{log(g)}$ & 4.53 & 0.10 & SPC \citep{buchhave+2012} & - \\
\hline
$\rm{vsin\textit{i} (km/s)}$ & 1.0 & 1.0 & SpecMatch-Synthetic &  Y \\
$\rm{vsin\textit{i} (km/s)}$ & 1.4 & 0.5 & SPC \citep{buchhave+2012} & - \\
\hline
$\rm{P_{rot} (d)}$ & 47 & 3 & Calculated using \citet{Mamajek:2008} & - \\
$\rm{P_{rot}/sin(i) (d)}$ & 42 & 40 & Calculated using $\rm{vsin\textit{i}}$ and $\rm{R_*}$ & - \\
$\rm{P_{rot} (d)}$ & 38 & $\rm{{}^{+7}_{-14}}$ & Calculated using \citet{claytor+2020} & Y\\
\hline
$\rm{{[Fe/H]}~(dex)}$ & 0.17 & 0.06 & SpecMatch-Synthetic \citep{specmatch-synth} & Y \\
$\rm{{[M/H]}~(dex)}$ & 0.17 & 0.08 & SPC \citep{buchhave+2012} & Y \\
\hline 
\enddata
\end{deluxetable*}

\section{Data Analysis} \label{sec:analysis}
\subsection{Photometric Fit} \label{sec:phot}
We first performed an MCMC fit of the \tess{} photometry using the \textit{emcee} package \citep{emcee} for each of the four transiting planets. We used the planet radii, periods, and transit times reported by the TOI catalog \citep{guerrero+2021} to initiate \texttt{batman} \citep{batman} transit models. We set planet eccentricities and arguments of periastron to 0 for simplicity, and used Kepler's Third Law to calculate an initial guess for the semi-major axes of the planets. Based on the effective temperature, $\log(g)$, and metallicity of TOI-1246, we adopt test values of $\rm{u_1 = 0.1}$ and $\rm{u_2 = 0.4}$ for quandratic limb darkening parameters following \citet{claret2017}. We used 48 walkers, and varied the following parameters for each planet: transit time, orbital period, planet radius, semi-major axis, and inclination. We assessed convergence using the integrated auto-correlation time.

Using the results of the initial MCMC analysis and using the stellar parameters derived in Section \ref{sec:star} as priors, we then performed a joint photometric fit of the four planets using the $\textsf{exoplanet}$ package \citep{exoplanet:exoplanet}. The photometric fit included a total of 21 free parameters. Each planet had four free parameters: orbital period ($P$), epoch ($T_0$), planet radius relative to stellar radius ($R_p/R_*$), and impact parameter ($b$). There are five further global parameters: quadratic limb darkening parameters ($u_1, u_2$), stellar mass ($M_*$) and radius ($R_*$), and a mean flux value ($\mu$). We assumed planet eccentricities to be 0 (motivated in Section \ref{sec:RV}). We chose to use epochs roughly halfway through the overall photometric baseline in order to speed up the photometric fit and reduce rounding errors. We calculated the transit times by assuming a linear emphemeris and propagating the transit midpoint time forwards to near 2459000 $\rm{BJD_{TDB}}$. The resulting planet radii are consistent with the radii found by the SPOC pipeline. 

\begin{figure*}[t]
    \includegraphics[width=\textwidth]{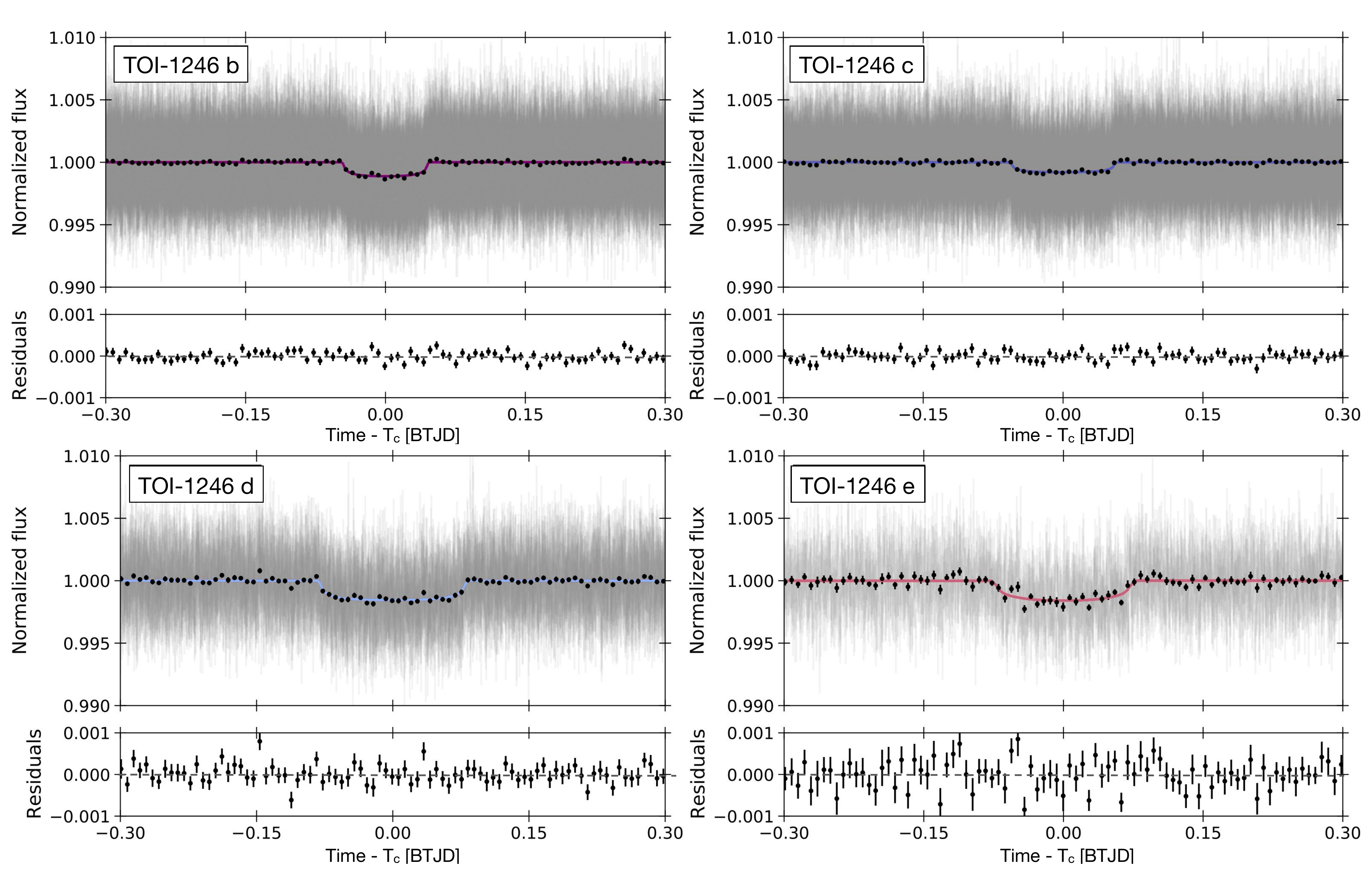}
    \centering
    \caption{\tess{} light curves phased at the orbital periods of the four transiting planets. Grey data are individual data points, black data points are binned data, and the overlaid curves are our best fit GP and TTV model (TOI-1246 b: purple, TOI-1246 c: blue, TOI-1246 d: light blue, TOI-1246 e: pink). Residuals are shown beneath each plot.}
    \label{fig:phasedtransits}
\end{figure*}

However, individual analysis of the system's transits, in particular those of the outermost planet 1246 e, showed some notable discrepancies with our transit models. There was non-negligible structure in the out-of-transit flux baseline, and the modelled ingresses and egresses of TOI-1246 e and  d were somewhat discrepant from the data. We attempted to ``whiten'' the photometry, and smooth out short time-scale stellar activity by applying a Gaussian Process (GP) in addition to the transit fits \citep{foreman-mackey+2017}. We used the Simple Harmonic Oscillator (SHO) kernel built into the \textsf{exoplanet} package \citep{exoplanet:exoplanet}. We varied two hyper-parameters: $\rho$ with a log normal prior with $\mu_\rho$ = 0 and $\sigma_\rho$ = 10, and $\sigma$ with a log normal prior with a mean equal to the standard deviation of the light curve flux, and a standard deviation of 10. We hold the final hyper-parameter fixed with a value of $\rm{Q = \frac{1}{\sqrt{2}}}$. While there was some modest improvement, much of the structure remained after such a fit. 

We therefore concluded that the poor fit of our model for TOI-1246 d and e was due to unmodelled TTVs. Therefore, we performed a TTV analysis with the Python Tool for Transit Variations (PyTTV) that models the transits using \texttt{Pytransit} \citep{Parviainen2015} and the stellar variability as a GP with a matern 3/2 kernel using \texttt{celerite} \citep{Foreman-Mackey2017}. This fit included the following free parameters: orbital period ($P$), epoch ($_0$), planet radius relative to stellar radius ($R_p/R_*$), transit midpoints ($t_c$) and impact parameter ($b$) for all four planets, while stellar density, and the quadratic limb darkening parameters ($u_1, u_2$) were shared for all the planets. The transits from all the planets in the TOI-1246 system were fit jointly by modeling them with the quadratic \citet{2002MandelAgol} transit model implemented in Pytransit via a Taylor-series expansion \citep{ParviainenKorth2020}. Section \ref{sec:ttvs} describes the measured TTVs in more detail. We estimated the model parameter posteriors using MCMC-sampling (emcee; \citealt{emcee}), and find that the transit midpoints show variations from a linear ephemeris. We show the phase-folded transits accounting for the TTVs in Figure \ref{fig:phasedtransits}, and calculate the following radii: \mbox{$\rm{R_{b} =~2.97 \pm 0.06~R_\oplus}$}, \mbox{$\rm{ R_{c} =~2.47 \pm 0.08~R_\oplus}$}, \mbox{$\rm{R_{d} =~3.46 \pm 0.09~R_\oplus}$}, and  \mbox{$\rm{R_{e} =~3.72 \pm 0.16~R_\oplus}$}. We also derive a stellar density of 2.47 $\rm{\pm 0.25 g/cm^3}$. This corresponds to a stellar mass of 1.12 $\rm{\pm 0.16~M_\odot}$, which is consistent with values in Table \ref{tab:stellar}. We apply the radius correction factor of 1.015 calculated in Section \ref{sec:ao} and report the final planet radii and other system parameters in Table \ref{tab:planet}.

Alongside the fits described above, we also fit the \tess{} photometry using the \texttt{TATER} code (Harada et al., in prep). \texttt{TATER} (the Tess trAnsiT fittER) is a custom Python tool which applies the \texttt{Transit Least Squares} \citep[TLS;][]{hippke+2019} algorithm to \tess{} photometry accessed via the Lightkurve package \citep{lightkurve} to iteratively search for transit signals, and implements an MCMC sampler to determine a best-fit \texttt{batman} transit model \citep{batman}  for each significant detection. \texttt{TATER} successfully recovered four transit signals consistent with the preferred fits for TOI-1246 b, TOI-1246 c, TOI-1246 d, and TOI-1246 e. It did not identify any significant transit signals at periods longer than that of TOI-1246 e and shorter than $\sim250$ d. The current photometric baseline of \tess{} observations is 752 d, with a large data gap, so a planet with an orbital period $\lesssim 250$ d could transit three times in the light curve. This suggests that the system lacks additional longer-period transiting planets with similar radii to the known transiting planets. 

\begin{deluxetable*}{lcccccccccc}
\tablecaption{TOI-1246 System Parameters \label{tab:planet}}
\rotate
\tablehead{\multicolumn{1}{l}{\textbf{Parameter}} & 
\multicolumn{1}{c}{\textbf{Value}} &
\multicolumn{2}{c}{\textbf{TOI-1246 b}} &
\multicolumn{2}{c}{\textbf{TOI-1246 c}} &
\multicolumn{2}{c}{\textbf{TOI-1246 d}} &
\multicolumn{2}{c}{\textbf{TOI-1246 e}} &  \multicolumn{1}{c}{\textbf{Source}} \\ & 
& Value & Error & Value & Error & Value & Error & Value & Error & }
\startdata
\textbf{Photometric Parameters} &&&&&&&&&& \\
Period (days)&& 4.30744 & 0.00002 & 5.904144 & 0.000083 & 18.65590 & 0.00048 & 37.9216
& 0.0010 & Photometric Fit\\
Epoch-2457000 (BTJD) && 1686.5658
& 0.0010 & 1683.4661 & 0.0027 & 1688.9653 & 0.0090 & 1700.7134 & 0.0089 & Photometric Fit\\
Impact parameter, b && 0.49 & 0.07 & 0.20 & 0.15 & 0.29 & 0.11 & 0.73 & 0.04 & Photometric Fit\\
Equilibrium Temperature\tablenotemark{a} (K) && 955 && 860 && 586 && 462 && ExoFOP-TESS\\
$\rm{R_p/R_*}$&& 0.031 & 0.001 & 0.026 & 0.001 & 0.036 & 0.001 & 0.039 & 0.002 & Photometric Fit\\
Radius ($R_\oplus$) && 3.01 & 0.06 & 2.51 & 0.08 & 3.51 & 0.09 & 3.78 & 0.16  & From $\rm{R_p/R_*}$\\
Semi-major axis, a ($\rm{R_*}$) && 13.4 & 0.5 & 16.6 & 0.6 & 35.7 & 1.2 & 57.3 & 2.0 & Photometric Fit\\
Semi-major axis, a (AU) && 0.049 & 0.002 & 0.061 & 0.002 & 0.131 & 0.004 & 0.211 & 0.007 & Photometric Fit\\
Inclination (deg) && 87.9 & 0.4 & 89.3 & 0.5 & 89.5 & 0.2 & 89.3 & 0.1 & From a  ($\rm{R_*}$), b\\
Limb darkening, $\rm{u_1}$ & 0.12 $\pm$ 0.10  &&&&&&&&&Photometric Fit\\
Limb darkening, $\rm{u_2}$ & 0.70 $\pm$ 0.17   &&&&&&&&&Photometric Fit\\
\hline
\textbf{Spectroscopic Parameters}&&&&&&&&&& \\
RV Semi-Amplitude (m/s) && 3.44 & 0.64 & 2.99 & 0.62 & 1.66 & 0.64 & 3.77 & 0.65 & \texttt{Radvel}\\
HIRES RV Offset (m/s) & 2.08 $\pm$ 0.70    &&&&&&&&& \texttt{Radvel} \\
HARPS-N RV Offset (m/s) & 4.87 $\pm$ 0.85    &&&&&&&&& \texttt{Radvel} \\
HIRES RV Jitter (m/s) & 2.88 $\pm$ 0.34    &&&&&&&&& \texttt{Radvel} \\
HARPS-N RV Jitter (m/s) & 3.18 $\pm$ 0.60    &&&&&&&&& \texttt{Radvel}l \\
Planet mass ($M_\oplus$) && 8.1 & 1.1 & 8.8 & 1.2 & 5.3 & 1.7 & 14.8 & 2.3 & \texttt{Radvel}\\
Density ($g/cm^3$) && 1.63 & 0.23 & 3.21 & 0.44 & 0.70 & 0.23 & 1.51 & 0.26 & From $\rm{M_p}$, $\rm{R_p}$\\
\hline
\textbf{Atmospheric Parameters} &&&&&&&&&& \\
Insolation Flux ($\rm{F_\oplus}$) && 196 & & 129 & & 28 & & 11 & & ExoFOP-TESS \\
Transmission Spectroscopy Metric && 48.9 & $\rm{{}^{+9.7}_{-7.6}}$ & 22.9 & $\rm{{}^{+5.0}_{-3.9}}$ & 72.7 & $\rm{{}^{+35.9}_{-19.2}}$ & 24.5 & $\rm{{}^{+6.3}_{-4.9}}$& Using \citet{kempton+2018}\\
\enddata
\tablenotetext{a}{Assuming a Bond albedo of 0.3.}
\end{deluxetable*}

\subsection{Anticipated $\&$ Observed Transit Timing Variations} \label{sec:ttvs}
A subset of transiting planets in multi-planet systems are amenable to mass measurements using the TTV method \citep{hadden+2017, agol+fabrycky2018}. TTVs in systems with only one known transiting planet have also been used to detect additional planets that were not initially detected \citep[e.g.][]{ballard+2011,lam+2020}. The NASA Exoplanet Archive reports that at least 325 of the 4531 confirmed planets display TTVs (as of 5 November 2021).  The \tess{} mission was predicted to find $\rm{\sim90}$ systems that exhibit TTVs during its prime and extended mission \citep{hadden+2019}. The NASA Exoplanet Archive reports that five \tess{} confirmed planets exhibit TTVs (as of 5 November 2021).

TOI-1246 d (P = 18.66 d) and TOI-1246 e (P = 37.92 d) lie just exterior to the 2:1 mean motion resonance, with a period ratio of 2.03. We predicted the magnitude of  TTV signals for these planets using the equations laid out in \citet{lithwick+2012}. We used the planet masses and periods reported in Table \ref{tab:planet}, and assumed circular orbits. The estimated magnitude of the TTVs in this system \citep{lithwick+2012} are 22 minutes for TOI-1246 d (the inner planet of the pair), and 4 minutes for TOI-1246 e. We also calculated the TTV super-period to be \mbox{$P_{super}=$ 1264 d}. The photometric data we analyzed cover 384 days within a time span of 632 days, and so only cover $50\%$ of the super-period. However, an additional 273 days of data will be collected in Cycle 4, leading to an overall baseline of 1141 days. This will increase the coverage of the super-period to around $90\%$. The upcoming sectors will observe an additional 54 transits of TOI-1246 b, 40 transits of TOI-1246 c, 13 transits of TOI-1246 d, and 6 transits of TOI-1246 e. These observations will double the number of observed transits for TOI-1246 e.

\begin{figure*}
    \includegraphics[width=\textwidth]{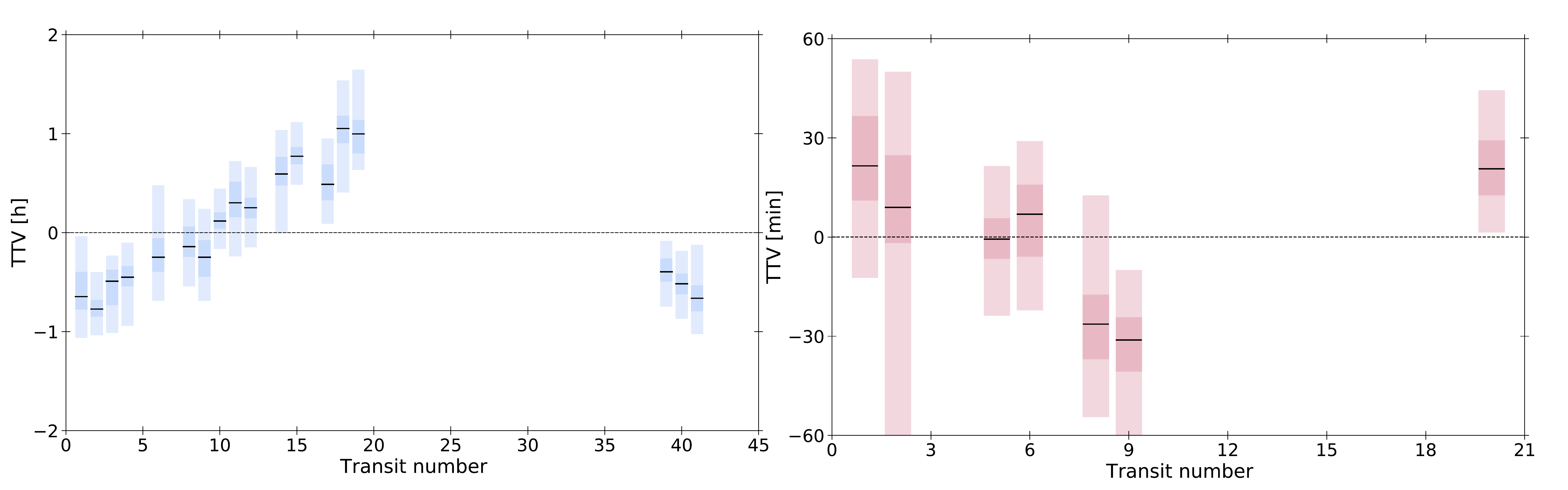}
    \centering
    \caption{Observed - Computed plots for TOI-1246 d (left) and TOI-1246 e (right). The shaded area of each bar represents the one- and three-sigma standard deviations of each TTV measurement. Both planets show strong evidence of TTVs. Transit number 2 of TOI-1246 e was not fully observed by \tess{}, leading to a larger error bar on the TTV measurement.}
    \label{fig:ttvs}
\end{figure*}

As described in Section \ref{sec:phot}, we fit the individual transit midpoints for TOI-1246 d and TOI-1246 e in order to measure TTVs. Figure \ref{fig:ttvs} shows the difference between the observed and calculated transit midpoints. We see clear evidence of TTVs for both planets, and note that they are larger in amplitude than the predicted magnitudes derived from \citet{lithwick+2012}.This may be due to the effects of non-zero orbital eccentricities or the effect of a fifth planet candidate in the system (see Section \ref{sec:fifth_in_RVs}) which may form a resonant chain with TOI-1246 d and TOI-1246 e. We do not investigate whether TOI-1246 d and TOI-1246 e are in resonance in this work. We will explore this result further in a follow-up work using the upcoming \tess{} photometry.

\subsection{Radial Velocity Analysis} \label{sec:RV}
\begin{figure*}
    \includegraphics[width=\textwidth]{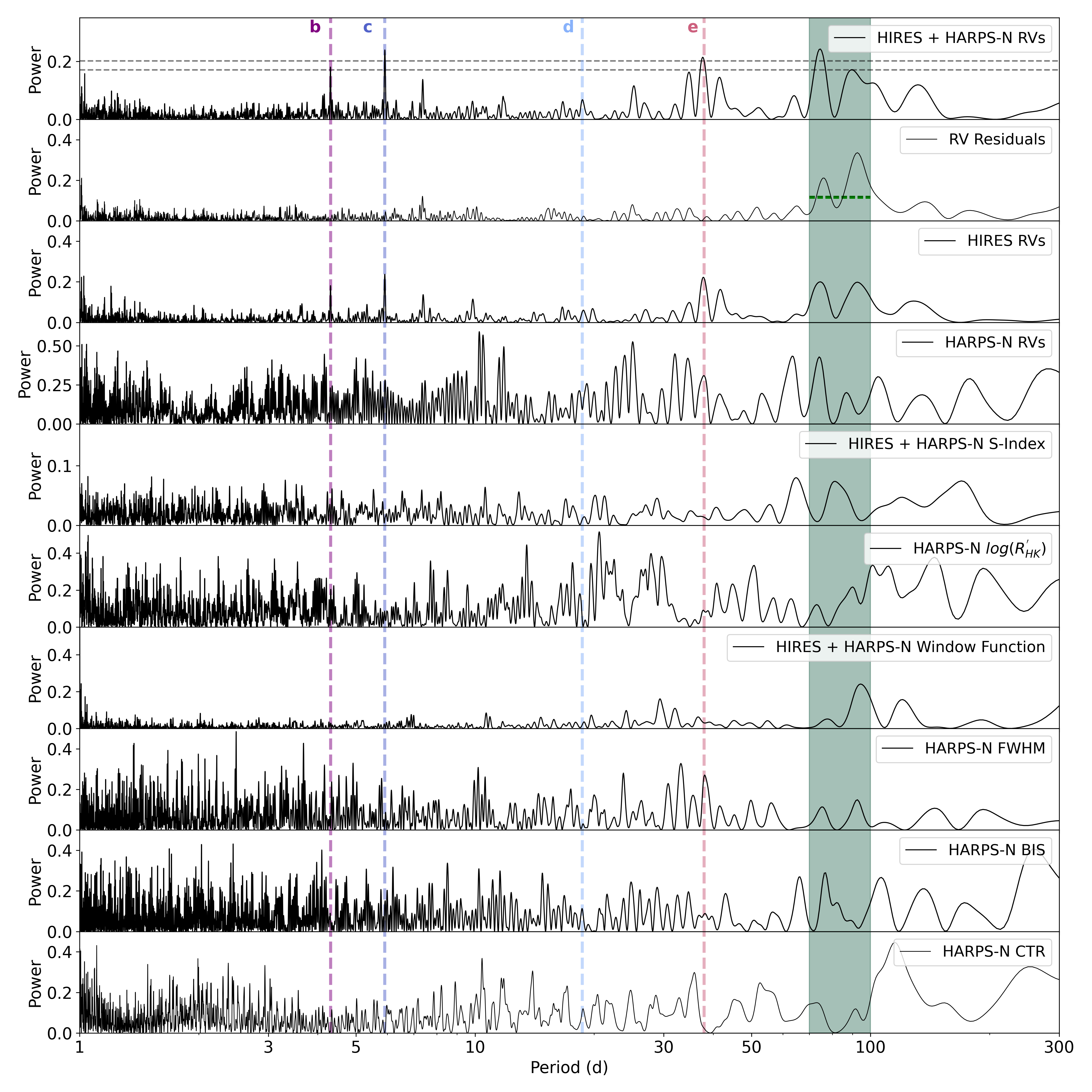}
    \centering
    \caption{Lomb-Scargle periodograms for the RV data, RV residuals from a four-planet fit, activity indicators, and spectral window function for TOI-1246 using Keck/HIRES and TNG/HARPS-N data (see legends). The vertical colored lines indicate the periods of the four transiting planets in the system (TOI-1246 b: purple, TOI-1246 c: blue, TOI-1246 d: light blue, TOI-1246 e: pink). The green region indicates the range of likely periods for the fifth candidate planet. We consider the peaks at both 76 and 94 d, and choose to report the fifth candidate planet at 94 d in this work. In the top panel, the horizontal dashed lines represent the $1\%$ and $0.1\%$ false-alarm levels. In the second panel, we show two closely spaced horizontal dashed lines in the range of period space we consider. These lines represent the 99.99th percentile of 10,000 bootstrap samples of these residuals, in dark green (for the 76 d signal) and green (for the 94 d signal), and show that both of these signals are statistically significant.}
    \label{fig:ls_periodograms_combo}
\end{figure*}

We performed a fit of the precision radial velocity data collected using Keck/HIRES and TNG/HARPS-N using the \texttt{RadVel} package \citep{radvel}. Figure \ref{fig:ls_periodograms_combo} shows the Lomb-Scargle \citep{lomb1976, scargle1982} periodograms for the RV data. 

In our RV analysis, we fixed the linear-ephemeris orbital periods and planet transit times to the values measured from \tess{} photometry. We restricted the semi-amplitude of each planet ($K$) to be positive, in order to constrain planet masses > 0, which is physically motivated. For each instrument, we introduce an RV offset term ($\gamma$), and an RV jitter term ($\sigma$), for a total of 4 instrumental parameters. We also considered an RV trend term ($\dot{\gamma}$), although our fits preferred values near zero for this parameter. In our simplest model, we assumed circular orbits for the four transiting planets, resulting in 9 free parameters: 4 semi-amplitudes, 2 jitter parameters, 2 RV offset parameters and 1 RV trend parameter.

We performed several different fits with a variety of eccentricity priors. The most restrictive eccentricity prior used the orbit-crossing eccentricity of each planet (TOI-1246 d: $\rm{e_{cross} = 0.536}$, TOI-1246 b: $\rm{e_{cross} = 0.234}$, TOI-1246 c: $\rm{e_{cross} = 0.1897}$, and TOI-1246 e: $\rm{e_{cross} = 0.377}$) as an upper limit for eccentricity. The orbit-crossing eccentricity is the eccentricity at which the orbit of a planet in the system would cross that of an adjacent planet (assuming the adjacent planet's orbit is circular). In less restrictive cases, we explored letting eccentricities vary fully from $e=0$ to $e=1$ and letting eccentricities vary up to upper limits set by our analysis of system stability. We also performed fits in which the three inner planets had circular orbits and the eccentricity of TOI-1246 e was allowed to vary. Regardless of the specific choice of eccentricity prior, we found that the resulting planet mass estimates remained within the $~1\sigma$ confidence interval of the circular orbit solutions. We report masses from the circular orbit solutions as motivated below and in Section \ref{sec:stability}. Figure \ref{fig:rv_fit} shows both the corresponding global fit and the phased RV model with other planet models subtracted for each planet.

Our decision to adopt the zero-eccentricity fit results is supported by \citet{yee+2021}, who investigated a sample of 19 \kep{} compact multi-planet systems with precise planet masses and eccentricities measured using TTVs. They found that the planet eccentricities in these systems were significantly (from a few to 10 times) smaller than required for dynamic stability, and even smaller than the orbit crossing limits. In other words, these systems are over-stable, implying that planet eccentricities must be damped if these planets initially formed through giant impacts. TOI-1246 has two planets near resonance which exhibit TTVs, and is therefore similar to the planet sample considered by \citet{yee+2021}. \citet{vaneylen+2015} and \citet{vaneylen+19} further find that planets in \textit{Kepler} multi-planet systems tend to have smaller eccentricities than and have a distinct eccentricity distribution to planets in systems with only one transiting planet detected. These results strengthen our choice to report the results of the zero-eccentricity RV fits. 

\subsection{A Fifth Periodic Signal in the RV Data}\label{sec:fifth_in_RVs}
\begin{deluxetable*}{ccccc}
\label{tab:bic}
\tablehead{\colhead{$\rm{N_{planets}}$} & \colhead{Additional Planet Period(s)} & \colhead{Additional Free Parameters} & \colhead{$\rm{N_{free}}$} & \colhead{BIC}}
\startdata
4 & - & - & 7 & 741.21 \\ 
5 & 76.2 d & $P_f, t_{c,f}, K_f$ & 10 & 708.09 \\ 
5 & 93.8 d & $P_f, t_{c,f}, K_f$ & 10 & 704.53\\ 
6 & 76.0 d and 94.3 d & $P_f, t_{c,f}, P_g, t_{c,g}, K_f, K_g$ & 13 & 702.96
\enddata
\caption{Selected results of RV analyses to compare the Bayesian Inference Criterion (BIC) for models with varying planet number ($\rm{N_{planets}}$), and additional planet candidate period(s). All models include the following free parameters: $K_b, K_c, K_d, K_e, \gamma, \dot{\gamma}$. We set the RV jitter terms to constant values determined by an initial five-planet fit in order to reduce the complexity of the model comparison. The five-planet models are preferred relative to the four-planet model.}
\end{deluxetable*}

An additional consideration in our analysis is a fifth periodic signal in the RV data with a period longer than that of TOI-1246 e. The Lomb-Scargle periodogram of the RV data residuals from a four-planet model includes signals at both 76.2 d and 93.8 d (see Figure \ref{fig:ls_periodograms_combo}). We also computed an $\rm{l_1}$ periodogram \citep{hara+2017f} of the RVs. The $\rm{l_1}$ periodogram was developed to search for periodic signals in RV data, and aims to reduce the number of peaks due to aliasing when compared to a Lomb-Scargle periodogram. We find evidence of signals at $\rm{\sim74d}$ and $\rm{\sim94d}$ in the $\rm{l_1}$ periodogram. To test the robustness of these signals, we performed bootstrap sampling \citep{efron1979} on the residuals from a four-planet RV fit. The signals are robust to bootstrap sampling to $>99.99\%$ confidence, and thus must be considered in our analysis.

\subsection{Possible Explanations for the Fifth Signal}
It is possible that this fifth Keplerian signal is a result of non-zero eccentricity of TOI-1246 e that is not addressed in our zero-eccentricity RV model. In order to test this possibility, we fit the RV data set with a four-planet model and allow the eccentricity of TOI-1246 e to vary. This model reported an eccentricity of $\rm{0.1~\pm~0.1}$ for TOI-1246 e, and does not remove the additional signals seen in the RV residuals between 70 and 100 d. Both the 76 d and 94 d signals remain robust to $>99.99\%$ after bootstrap sampling the RV residuals of this fit. Therefore, we do not consider unmodelled eccentricity as a plausible explanation for this signal.

Another possible explanation is that it is related to stellar activity and rotation. Figure \ref{fig:ls_periodograms_combo} shows that there is a peak in the S-Index Lomb-Scargle periodogram at ~80 d, and a peak at ~76 d in the HARPS-N BIS periodogram. However, our analysis of the photometric light curve and predictions for the stellar rotation period are inconsistent with a rotation-related signal at either of the two candidate periods with with $> 99.9$\% confidence (more details in Section \ref{sec:star}). Therefore, we do not believe this signal to be related to star spots modulated by stellar rotation. Furthermore, there are no significant peaks between 70 and 100 d in the CCF\_CTR or FWHM spectral activity indicator periodograms derived from HARPS-N data. The signal at $\rm{80~d}$ in the S-Index periodogram may arise from the window function (and thus not be astrophysical) or from stellar activity on the timescale of 80 days that our analysis has not sufficiently considered.

Finally, we consider the potential sampling effect of the RV observations. The spectral window function (shown in Figure \ref{fig:ls_periodograms_combo}) shows a peak near 90 days, and may be clouding our picture of signals in the 70-100 day range. Furthermore, both the 76.2 d and 93.8 d candidate periods for the fifth signal are near harmonics of other planet periods (TOI-1246 e: $37.92 {\rm d} \approx 76{\rm d}/2$, and TOI-1246 d: $18.66 {\rm d} \approx 94{\rm d}/5$, respectively), and so may appear more significant in the RV periodogram than they actually are due to aliasing. We are collecting additional RVs using WIYN/NEID in the 2022A observing semester in order to mitigate these aliasing effects resulting from uneven sampling \citep{DawsonFabrycky2010}. WIYN is less susceptible to the 95-day seasonal window function than Keck (where we collected \HIRESRVs{} of \totalRVs{} RVs), and these RVs will be useful in ascertaining the period, validity, and nature of the long-period signals identified here.

Given that stellar rotation and un-modelled eccentricity are unlikely to be responsible for these long period signals, and that the signals are robust to bootstrap sampling, we tentatively attribute the signal to a fifth planet candidate with a period of either 76.2 d or 93.8 d. We refer to this planet candidate as planet f for the remainder of this paper, but emphasize that this does not imply confirmation of this candidate planet.

RV signals from non-transiting planets are expected to be common in systems with transiting planets \citep{he+2021}. The correlated RV signals from these unmodelled planets (or other unmodelled signals) can affect the accuracy of mass measurements derived from RV observations \citep[e.g.][]{bonfils+2018, cloutier+2017}. With this in mind, we inspected the extent to which our choice of underlying planet model affected the masses of the four transiting planets. We considered models with 4, 5, or 6 planets (i.e. the four transiting planets with 0, 1, or 2 additional outer planets), and with or without varying eccentricity, and found that the best-fit values for the planets masses were consistent to $1\sigma$. Thus, the uncertainties in the planet masses appear to be dominated by measurement uncertainties, rather than model selection. 

We also do not find any un-accounted-for transit events in the \tess{} photometry (see Section \ref{sec:phot}), implying that this fifth long-period planet candidate is likely non-transiting if similar in size to TOI-1246 e. We simulated light curves for TOI-1246 using the continuum flux in the original light curve and injected planets with a range of radii in order to assess our ability to recover planet transits of various depths. We are able to identify the individual transits of planets $\rm{\gtrsim 4~R_\oplus}$ by eye and can recover these signals. Given the minimum mass of $\sim25 M_\oplus$ associated with the fifth planet candidate, \citet{chen+kipping2017} predict a radius of  $\sim5~\rm{R_\oplus}$. As a result, we would expect to see evidence of a Neptune-sized long-period fifth planet candidate if it did transit. 

\subsection{Selecting a preferred model} \label{sec:model_comparison}

In order to determine which model to select as our preferred fit, we used the Bayesian Inference Criterion (BIC) to compare models (see Table \ref{tab:bic}). We interpret a $\Delta$BIC $> 5$ to indicate a model is preferred. We find that the five-planet models both have lower BICs than the four-planet model by $>5$, and thus are preferred. The five-planet model with $\rm{P_f} = 93.8$ d has a BIC that is slightly lower than that for the analogous model with $\rm{P_f} = 76.2$ d, and so we propose a candidate period of 93.8 d. We also considered the BIC for a six-planet model, with two additional planet candidates at 76 d and 94 d. The resulting BIC is smaller than those for the two five-planet models, but since the difference is small and we are hesitant to propose two additional planet candidates in this region of period space, we report the results of the 93.8 d five-planet model. 

\begin{figure*}
    \centering
    \includegraphics[width=\textwidth]{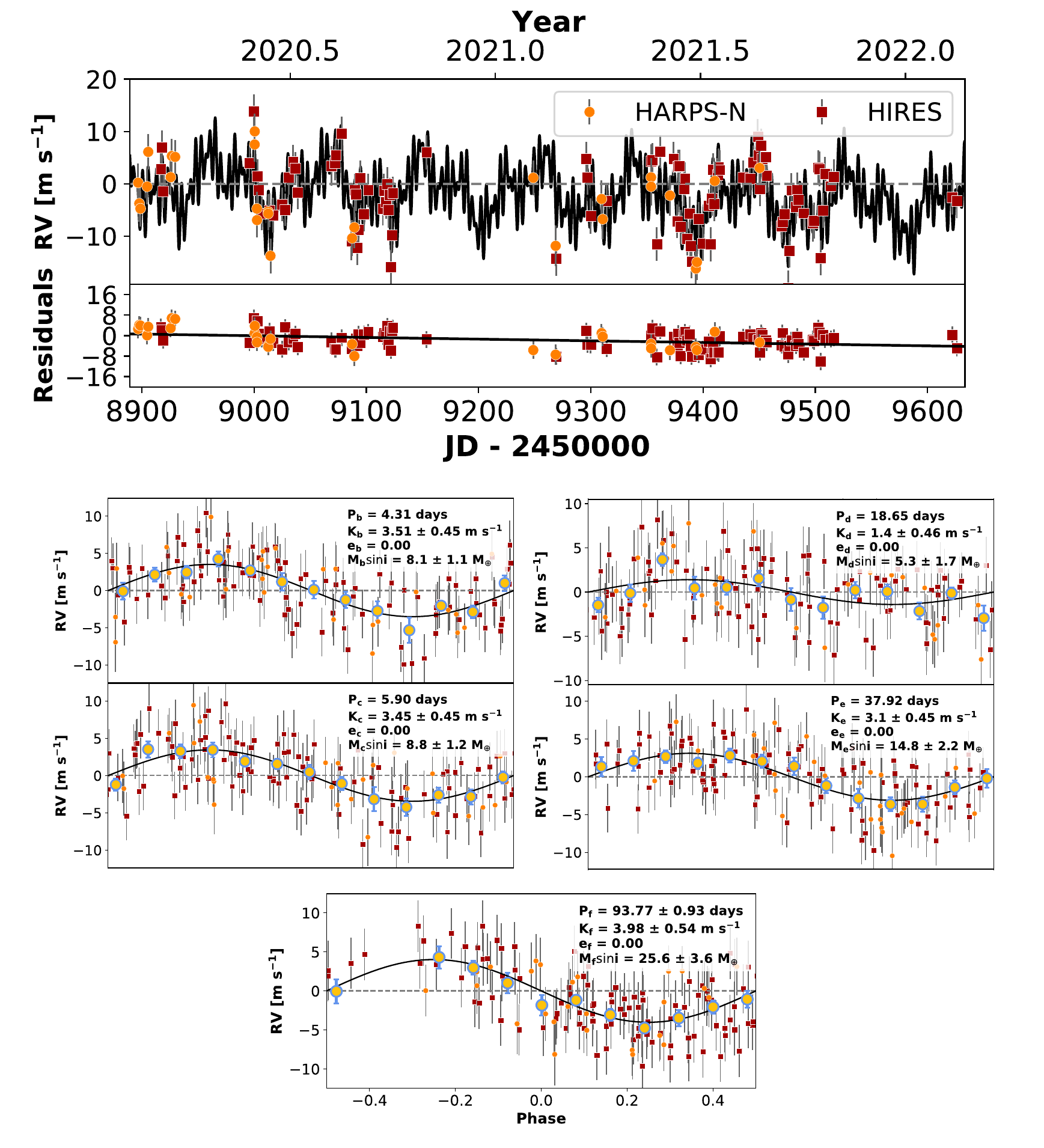}
    \caption{RV data and models for the four transiting planets and a fifth non-transiting planet candidate. The top panel shows the complete RV time series (orange points are TNG/HARPS-N data, and red points are Keck/HIRES) with the preferred five-planet RV model overlaid in black and with residuals (RMS = 4.3 m/s) shown below. The five panels beneath this show the RV data phased to each planets' orbital period (with binned data points in yellow and blue), with the other planets' signals removed, and with the RV model overlaid in black.}
    \label{fig:rv_fit}
\end{figure*}

The minimum mass of the fifth planet candidate resulting from the preferred model is $M \sin i  = 25.6 \pm 3.6 \rm{~M_\oplus}$, which is much more massive than the four transiting planets and the solar system ice giants. We report the corresponding mass measurements for the transiting planets ($M_{b} =$ \massc $M_\oplus$, $M_{c} =$ \massd $M_\oplus$, $M_{d} =$ \massb $M_\oplus$, $M_{e} =$ \masse $M_\oplus$) in Table \ref{tab:planet} and show the results of this fit in Figure \ref{fig:rv_fit}. We discuss the implications of the fifth candidate planet on the system architecture in Section~\ref{sec:fifth}.

There is still some structure remaining in the RV residuals of the five-planet fit (see the top panel of Figure \ref{fig:rv_fit}). We performed additional RV fits allowing the eccentricities of one, two, or all of the outer three planets (TOI-1246 d, TOI-1246 e, and the fifth planet candidate) to vary. We find that this does not remove this structure, indicating there is remaining unmodelled signal in the RV data set. We performed bootstrap sampling on these five-planet model residuals, and found that the peak at 76 d is significant in the residuals to $>99.99\%$. This indicates that the 76~d period signal cannot be explained as an alias of the 93.8~d period. This signal could correspond to a sixth planet candidate with a period of 76 d, but more data is needed to investigate this tentative hypothesis. 

\subsection{Stability}\label{sec:stability}
We explored the stability of the system using the SPOCK package \citep{Tamayo+2020}. In addition to modeling the system configuration adopted in Section \ref{sec:model_comparison}, we investigated how the stability changed as we varied planet eccentricities and the number of planets within the system. We did not consider the effects of the proper-motion companion TIC 230127303 because of its wide separation with TOI-1246 ($\sim 750$ AU).

When considering eccentricity variations, we were inspired by past observational and theoretical work. \citet{ford+2008} predicted the \kep{} mission would probe the eccentricity distribution of transiting planets. Several works have investigated subsets of this sample, using transit duration statistics to infer eccentricity distributions \citep[e.g.][]{fabrycky+2014,xie+2016, shabram+2016}. \citet{Mills+2019} considered a large ($\sim1000$) sample of planets with high-precision RVs from the California-\kep{} Survey, and found that the eccentricity distribution is well-described by a truncated Rayleigh distribution, in agreement with the previous works.

SPOCK estimates the stability of a planetary system with given planet parameters, and generates a probability of orbit stability, i.e. the probability that a planet on a given orbit survives $10^9$ orbits. Each simulation is ended and classed as unstable if any planets' Hill spheres overlap (see \textit{Materials and Methods} in \citealt{Tamayo+2020} for details). Using SPOCK, we evaluated system stability by drawing 10,000 realizations of system parameters from the following distributions:
\begin{itemize}
    \setlength\itemsep{0pt}
    \item Eccentricity: truncated Rayleigh distribution with a mode of 0.0355 (as per \citealt{Mills+2019}). 
    \item Orbital Period: held constant at the values reported in Table \ref{tab:planet}).
    \item Stellar Mass: Normal distribution  with a mean of $0.87 M_{\odot}$ and a standard deviation of $0 c M_{\odot}$ (see Table \ref{tab:stellar}).
    \item Inclinations: fixed for the four transiting planets with values determined using the impact parameters and semi-major axes reported in Table \ref{tab:planet}. We used a uniform distribution in the range $[0, 0.1736]$ for $\cos(i)$ for the non-transiting fifth planet candidate. Systems with intrinsically higher multiplicities have lower mutual inclinations (discussed in \citealt{he+2020}), and so a fifth planet candidate will likely be well-aligned with the other planets in this compact multi-planet system. 
    \item Planet Mass: normal distribution truncated to prevent negative planet masses (this only reduces the parameter space investigated by $0.4\%$ for TOI-1246 d, and not at all for the other planets). The mean and standard deviation for each normal distribution was set by the preferred RV model described in Section \ref{sec:model_comparison} and reported in Table \ref{tab:planet}. We used the inclinations described above to calculate planet masses from the minimum masses derived from the RV data.
    \item Argument of periastron, $\omega$: uniform distribution in the range $[0,2\pi]$ for each planet
\end{itemize}
We find that $\rm{13\%}$ of trial systems have $\rm{>0.34}$ probability of persisting over $10^9$ orbits, where 0.34 is the cutoff between stability/instability reported in \citealt{Tamayo+2020}. We note that this low stability value is likely due to the destabilizing effect of TOI-1246 e's location near mean motion resonance with TOI-1246 d. We simulated several system architectures with distributions as described above except for TOI-1246 e's orbital period, which we varied such that it was no longer near resonance with TOI-1246 d. We find that such systems have substantially higher stability (>95\% of systems have a $\rm{>0.34}$ chance of persisting over $10^9$ orbits). We will investigate the dynamics of this system more closely in a follow-up work.

We also used SPOCK to simulate only the four transiting planets, in order to consider the scenario where the fifth longer-period RV signal is not planetary in nature. We find that the four-planet system architecture is similarly stable to the scenario including the fifth planet candidate at 93.8 days. Therefore, the presence of a fifth planet candidate does not necessarily destabilize the system. 

We also considered how additional undetected planets would affect system stability using SPOCK. The long baseline of both photometric and spectroscopic observations of TOI-1246 mean we are sensitive to detecting planets with a large range of orbital periods. Given that it would be unlikely to miss such a short-period transiting planet with a transit depth similar to that of TOI-1246 b across the long photometric baseline, we are confident that there are no further transiting sub-Neptunes interior to TOI-1246 b. However, the relatively large gaps in period space between TOI-1246 c and TOI-1246 d, and between TOI-1246 d and TOI-1246 e raise the question of whether another inner planet could be hiding in the system. We performed two tests to evaluate the likelihood of the system containing another planet interior to TOI-1246 e. In each test, we drew 10,000 samples from the same distributions as described above, and added an additional $\rm{1 M_\oplus}$ planet with a period of 10.5 d or 26.6 d and an inclination of 88$\rm{\degree}$ (mean of the inclinations of the other transiting planets). The choice of periods was motivated by \kep{} multi-planet statistics. \citet{weiss+2018} show that planets in multi-planet systems are evenly spaced in log-period space, and so we chose to test periods which are evenly spaced in log-period space between the known transiting planets in the TOI-1246 system. Using the mass-radius relation of \citet{wolfgang+2016} for sub-Neptunes, we estimate a corresponding radius of $< 1 \rm{R_\oplus}$ (and a corresponding transit depth of $< 114~ppm$). Based on the results of the transit injection and recovery test discussed in Section \ref{sec:fifth_in_RVs}, we find that a $\rm{1~R_\oplus}$ test planet could transit but not be detected in transit data. The eccentricity and argument of periastron of the additional planet were drawn from the same distributions as in the default test. We performed these tests both including and excluding the long-period planet candidate (P = 93.8 d) found in the RV data. All of these tests resulted in a less stable system (see Table \ref{tab:spock}), which supports the theory that there are no further planets in the gaps between the detected planets in the system. 

\begin{deluxetable*}{ccccc}
\label{tab:spock}
\tablehead{\colhead{Number of Planets} & \colhead{Planet Period(s)} & \colhead{Transiting Planets Eccentricity} & \colhead{Fraction of Stable Samples}}
\startdata
4 & & $\rm{e_i = 0}$ &  0.15\\
4 & & $\rm{e_i = TruncatedRayleigh(\sigma = 0.0355)}$ &  0.13\\
4 & & $\rm{e_i = 0}$, $\rm{e_{.04} = 0.1 \pm 0.1}$ & 0.86 \\
4 & & $\rm{e_i = TruncatedRayleigh(\sigma = 0.0355)}$, $\rm{e_{.04} = 0.1 \pm 0.1}$ & 0.11 \\
4 & $P_e$ = 26 d & $e_i = 0$ & 0.96 \\
4 & $P_e$ = 45 d & $e_i = 0$ & 0.97 \\
5 & $P_5$ = 76.2 d & $\rm{e_i = TruncatedRayleigh(\sigma = 0.0355)}$ &  0.08\\
5 & $P_5$ = 93.8 d & $\rm{e_i = TruncatedRayleigh(\sigma = 0.0355)}$ &  0.13\\
5 & $P_5$ = 10.5 d & $\rm{e_i = TruncatedRayleigh(\sigma = 0.0355)}$ & 0.06\\
5 & $P_5$ = 26.6 d &$\rm{e_i = TruncatedRayleigh(\sigma = 0.0355)}$ & 0.03\\
6 & $P_5$ = 10.5 d, $P_6$ = 94 d & $\rm{e_i = TruncatedRayleigh(\sigma = 0.0355)}$ & 0.06\\
6 & $P_5 = 26.6 d$, $P_6$ = 94 d & $\rm{e_i = TruncatedRayleigh(\sigma = 0.0355)}$ & 0.03\\
\enddata
\caption{Results of SPOCK analysis of system stability. For planet periods not specified in the table, we used the reported planet periods from Table \ref{tab:planet}. For each SPOCK model tested, this table reports the fraction of the 10,000 samples drawn that remain stable, i.e. that have a $\geq 34\%$ probability of persisting over $\rm{10^9}$ orbits.}
\end{deluxetable*}

\section{Results and Discussion} \label{sec:results+discussion}

The TOI-1246 system hosts four transiting planets, and we find no evidence of additional transiting planets in the \tess{} photometry. We measured masses for these four transiting planets, and have found a longer-period signal in the RV data that may be planetary in nature. We also find that the system's stability is not significantly affected by a fifth planet candidate at 93.8 d. We list the planet properties from this work in Table \ref{tab:planet}.

\citet{millholland_winn2021} find that high-metallicity stars ([Fe/H]$>$0.0) tend to host multiple sub-Neptunes which are less uniform in size, and TOI-1246 conforms to this statistical prediction. The inner three transiting planets have distinct radii within the sub-Neptune population; TOI-1246 c lies at the peak of the radius distribution, while TOI-1246 b resides on the `occurrence cliff' \citep{fulton+2017}. TOI-1246 d and TOI-1246 e are among the largest sub-Neptunes. 

\subsection{Planet Bulk Densities and Composition} \label{sec:comps}
\begin{figure*}
    \centering
    \includegraphics[width=\textwidth]{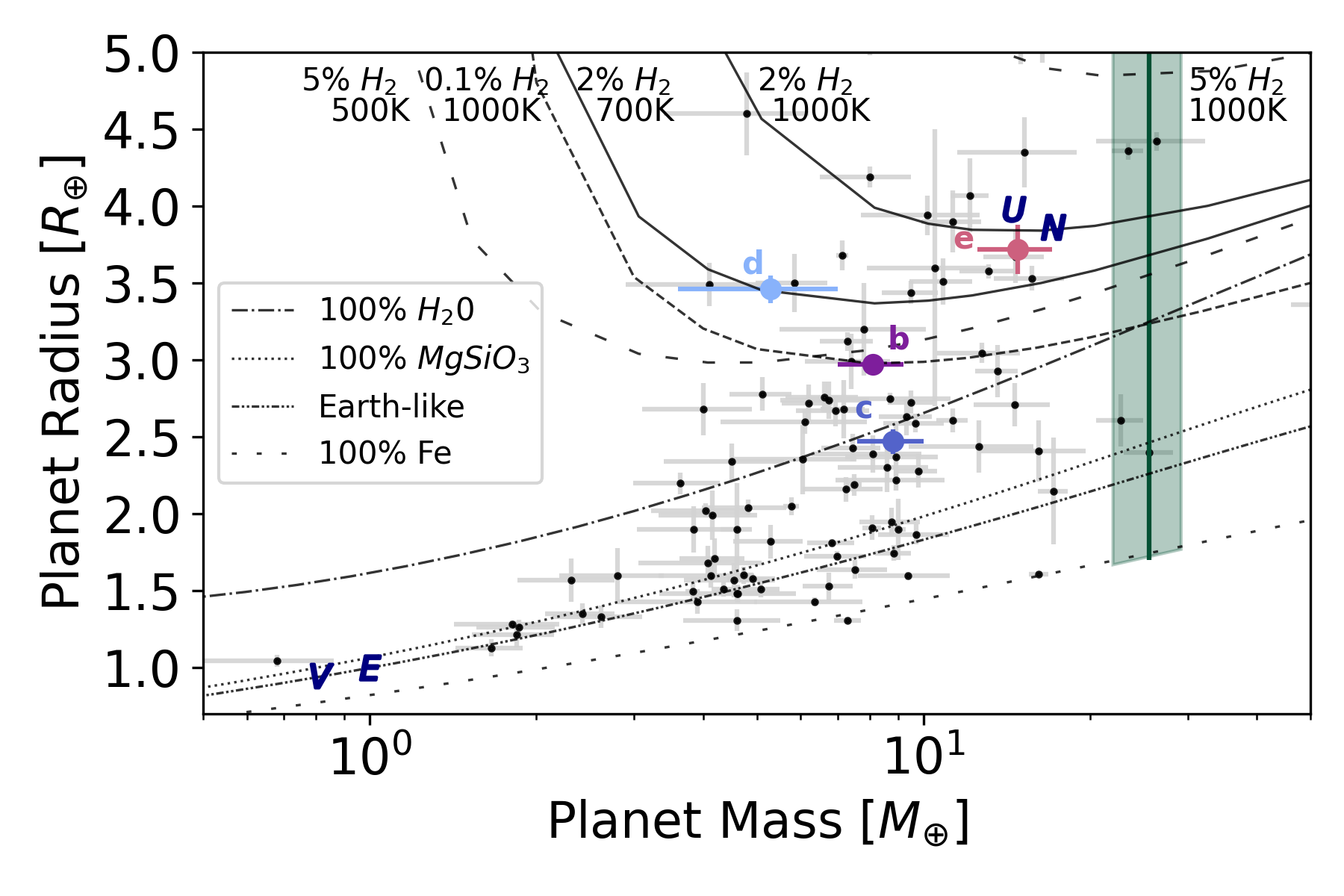}
    \caption{Mass-radius plot showing the four-planets orbiting TOI-1246 (TOI-1246 b: purple, TOI-1246 c: blue, TOI-1246 d: light blue, TOI-1246 e: pink). The parameter space that the fifth non-transiting planet candidate could occupy is shown in green (we exclude potential compositions more dense than $\rm{100\%~Fe}$). Theoretical composition curves from \citep{zeng+2019} are shown in grey. Earth (E), Venus (V), Uranus (U), and Neptune (N) are also shown for context, with the precise masses and radii for these planets lying in the center of the letter symbol. The subset of confirmed planets in systems with 2 or more planets, and with mass and radius measurements with $\rm{>3\sigma}$ precision are plotted in black with gray error bars.}
    \label{fig:massvsradius}
\end{figure*}

Figure \ref{fig:massvsradius} shows the TOI-1246 planets in mass-radius space. We find that the inner three planets in this system have similar masses, while the outermost planet is significantly more massive, with a mass similar to those of the ice giants in our Solar System. The fifth planet candidate has a minimum mass of  $25.6~\pm~3.6~\rm{M_\oplus}$, which is even more massive than the masses of Uranus and Neptune. TOI-1246 b and TOI-1246 c have the similar masses but distinct radii, and so comparing their atmospheres would be interesting for probing the formation and evolutionary history of the system. The four transiting planets also have quite varied densities: $\rho_{b} = 1.74 \pm 0.23 \rm{g~cm^{-3}}$, $\rho_{c} = 3.21 \pm 0.44 \rm{g~cm^{-3}}$, $\rho_{d} = 0.70 \pm 0.24 \rm{g~cm^{-3}}$, and $\rho_{e} = 1.62 \pm 0.25 \rm{g~cm^{-3}}$. All four transiting planets have densities lower than that of Earth\footnote{\url{https://nssdc.gsfc.nasa.gov/planetary/factsheet/earthfact.html}} ($\rho_\oplus = 5.514~\rm{g~cm^{-3}}$). TOI-1246 b and TOI-1246 e have similar bulk densities to that of Neptune\footnote{\url{https://nssdc.gsfc.nasa.gov/planetary/factsheet/neptunefact.html}} ($\rho_{\neptune} = 1.638~\rm{g~cm^{-3}}$). Although TOI-1246 d is a sub-Neptune, it has a similar bulk density to Saturn \footnote{\url{https://nssdc.gsfc.nasa.gov/planetary/factsheet/saturnfact.html}} ($\rho_{\saturn} = 0.67~\rm{g~cm^{-3}}$).

The TOI-1246 planets are located in well-populated regions of mass-radius parameter space. TOI-1246 b has a mass of \massc $~\rm{M_\oplus}$ and a radius of 2.97 $\rm{\pm}$ 0.06 $\rm{R_\oplus}$. It is consistent in mass and radius with Kepler-28 b \citep{steffen+2012}, Kepler-49 c \citep{steffen+2013,rowe+2014,jontof-hutter+2016}, Kepler-223 b \citep{rowe+2014,mills+2016}, Kepler-307 b \citep{xie2014}, and K2-266 d \citep{rodriguez+2018}. These planets are all members of compact multi-planet systems just as TOI-1246 b is. Furthermore, all of these planets exhibit TTVs, and the Kepler-223 system hosts a four-planet near-resonant chain. While TOI-1246 b does not orbit near or in resonance with another planet in the TOI-1246 system, it is interesting to note that all of these planets are in compact multi-planet systems and are either near resonance or have planetary siblings near resonance. 

TOI-1246 c (\massd $\rm{~M_\oplus}$, 2.47 $\rm{\pm}$ 0.08 $\rm{R_\oplus}$) is consistent in mass and radius with HD 5278 b, K2-138 d \citep{christiansen+2018}, HD 15337 c \citep{gandolfi+2019}, and K2-38 c \citep{sinukoff+2016}. HD 5278 b \citep{sozzetti+2021}, which has a similar insolation flux to that of TOI-1246 c ($\rm{S_{c} = 129 S_\oplus}$, $\rm{S_{HD 5278 b} = 132 S_\oplus}$), highlights how little can be inferred from a planet's bulk density. All of these planets lie near the $\rm{100\%~H_2O}$ composition line, and yet their masses and radii could be explained by a diversity of compositions. \citet{sozzetti+2021} suggest two options: a `wet' differentiated planet including a substantial water layer and a H/He gas envelope, and a `dry' planet with an iron core and silicate mantle. While \citet{sozzetti+2021} find that sub-Neptunes are expected to retain their volatile contents, the bulk density of TOI-1246 c is consistent with a variety of compositions. Atmospheric observations, such as transmission spectroscopy (see Section \ref{sec:atm}), are critical to distinguish between degenerate planet composition models \citep{rogers+seager2010}. 

TOI-1246 d (\massb $\rm{~M_\oplus}$, 3.46 $\rm{\pm}$ 0.09 $\rm{R_\oplus}$) has a mass and radius consistent with that of Kepler-11 d \citep{lissauer+2011}, Kepler-33 e \citep{lissauer+2012}, Kepler-79 e \citep{rowe+2014}, Kepler-177 b \citep{xie2014, jontof-hutter+2016, hadden+2017}, Kepler-223 c \citep{rowe+2014} and K2-32 b \citep{dai+2016}. Once again, all of these planets are in multi-planet systems, and all except Kepler-11 d exhibit TTVs, as does TOI-1246 d. Furthermore, the Kepler-79, Kepler-223 and K2-32 systems host multi-planet resonant chains of planets. If the fifth planet candidate in the TOI-1246 system is confirmed to be planetary, and if its period is found to be one of the candidate periods we suggest in this paper, TOI-1246 will also host a resonant chain of planets.

TOI-1246 e (\masse $\rm{~M_\oplus}$, 3.72 $\rm{\pm}$ 0.16 $\rm{R_\oplus}$) is consistent in mass and radius with GJ 3470 b, EPIC 249893012 c, WASP-47 d, Kepler-1661 b, and Kepler-595 b. Kepler-1661 b is a circumbinary planet which orbits near the hot edge of the habitable zone of a young (1-3 Gyr) binary system \citep{socia+2020}. EPIC 249893012 c \citep{hidalgo+2020}, on the other hand, orbits an evolved star with an age of $\rm{9.0^{+0.5}_{-0.6}}$ Gyr that is just leaving the main sequence. TOI-1246 is intermediate in age to these stars. WASP-47 d \citep{hellier+2012} and Kepler-595 b \citep{morton+2016, yoffe+2021}, like TOI-1246 e, are the outer planet in a pair near 2:1 resonance. They both show TTVs, as does GJ 3470 b \citep{bonfils+2012, awiphan+2016}. It is interesting to note that the majority of planets that are similar in mass and radius to the TOI-1246 planets also show TTVs, and that such planets are almost always found in multi-planet systems. 

Having compared the TOI-1246 planets to other planets with precisely known properties, we now consider possible atmospheric compositions for these worlds. The TOI-1246 planets are likely to harbor volatile envelopes that have been sculpted by their proximity to the host star and evolutionary history. We further investigate the potential compositions of these planets using the Exoplanet Composition Interpolator\footnote{\url{https://tools.emac.gsfc.nasa.gov/ECI/}}, which was developed using models from \citet{lopez+fortney2014}. These models assume an isothermal Earth-like 2:1 rock/iron core, a fully adiabatic interior for the H/He envelope, and a small isothermal radiative atmosphere atop the H/He envelope. This tool uses planet radii, masses, and insolation flux, as well as stellar age, to predict the core mass fraction and envelope mass fraction of each planet. The results of this interpolation indicate that TOI-1246 b has a $\sim5\%$ H/He envelope, while TOI-1246 c has a $\sim2\%$ H/He envelope, and TOI-1246 d and TOI-1246 e have $\sim10\%$ H/He envelopes by mass. Due to the uncertainty in the stellar age discussed in Section \ref{sec:star}, we evaluated these models for several stellar ages between 5 and 10 Gyr. We find that the results vary by $\sim1\%$, and the trend of envelope masses across the system does not change, so the envelope fractions reported are robust to larger uncertainties in stellar age than we have reported.

TOI-1246 b lies on the 0.1\% $\rm{H_2}$ composition curve for a 1000K planet. While TOI-1246 b has an equilibrium temperature of 950K, it is predicted to have a $\sim5\%$ H/He envelope by mass; and so lies far from its predicted location in mass-radius space to the top right of Figure \ref{fig:massvsradius}. TOI-1246 c lies near the $\rm{100\%~H_2O}$ composition curve, and is likely to have a higher density core (possibly with a substantial water fraction) surrounded by a low density envelope. TOI-1246 e has a similar size to both of the Solar System ice giants. Uranus and Neptune have similar internal structures; standard models include a rocky core surrounded by a thick massive icy mantle and a low-mass H/He atmosphere \citep{podolak+1995}. This may indicate a potential composition for TOI-1246 e, but due to inherent degeneracy in mass and radius measurements and the differences in $\rm{F_{insol}}$, we cannot conclusively determine the interior structure of TOI-1246 or the other TOI-1246 planets.

\subsection{System Architecture}

The dynamical architecture of TOI-1246 was previously probed by \citet{dietrich+2020}, who formalized a model (DYNAMITE\footnote{\url{https://github.com/JeremyDietrich/dynamite}}) to predict the periods, radii and inclinations of undetected planets in multi-planet systems using population statistics. They tested the DYNAMITE model on several \tess{} multi-planet systems, including TOI-1246. At the time of publication, TOI-1246 e was not yet a TOI, and so \citet{dietrich+2020} considered TOI-1246 as a three-planet system. They predict a fourth planet in the system with a period of $10.4^{+1.93}_{-1.94}~\rm{d}$, a radius of $2.91^{+0.921}_{-0.737}~\rm{R_\oplus}$, and a predicted transit probability of $0.91 \pm 0.001$ (Table 2 in \citealt{dietrich+2020}). If such a planet were transiting, we would expect to detect it in the \tess{} light curve, as it has a similar radius and period to other planets in the system. However, we do not find evidence of such a planet in the \tess{} light curve. Furthermore, extending our SPOCK analysis of system stability to include such a planet destabilizes the system (system stability is reduced by a factor of 2; see Table \ref{tab:spock}). While the predicted period does not match the actual fourth planet in the system (TOI-1246 e), it does lie evenly between TOI-1246 c and TOI-1246 d in log-period space, as predicted by \kep{} multi-planet statistics. There is some evidence in the RV data of a signal at around 10 d (see Figure \ref{fig:ls_periodograms_combo}), but we attribute it to the spectral window function of TNG/HARPS-N observations. Therefore, we find that there is insufficient evidence for a planet at 10 days, although such a low-mass planet in such an orbit is not ruled out by the data. \citet{dietrich+2020} do not indicate that a fifth planet with a period between 50 and 100 days would be likely, but they only consider the probability normalized to 1 injected planet, and so this does not preclude a fifth, exterior planet from being a likely addition to the TOI-1246 from a dynamical point of view.

\begin{figure*}
    \centering
    \includegraphics[width=11.5cm]{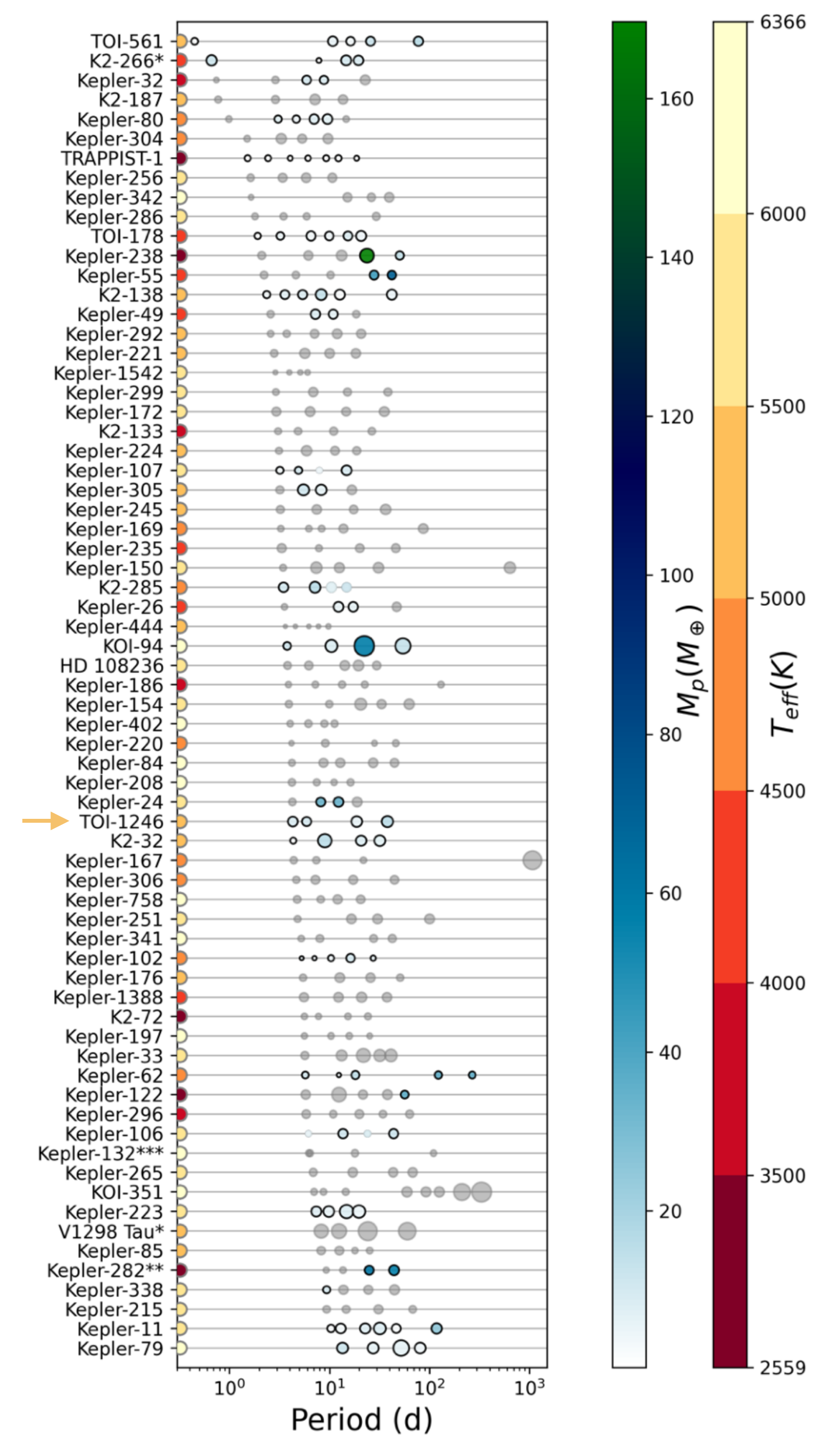}
    \caption{High-multiplicity ($\rm{N_{pl}} \geq 4$) planet systems where all planets are confirmed, ordered by increasing innermost orbital period. Point size is determined by planet radius, and planets with measured masses are shaded according to the colorbar on the right of the figure. Planets with upper limit mass measurements do not have a black edge, and are coloured according to the upper limit. The effective temperatures of the systems' host stars are indicated using the semi-circles on the left of each row, shaded according to the second colorbar. TOI-1246 is one of only six systems with measured masses for all four planets, and has a unique system architecture within this group. *V1298 Tau and K2-266 are the only systems in this population which are brighter than TOI-1246. **We note that the masses shown for the two outer planets in the Kepler-282 system are nominal TTV masses and are likely over-estimates \citep{xie2014}}. ***Kepler-132 is a binary star system that hosts 4 planets (at 6.2d, 6.4d, 18.0d and 110d), but it is unknown which planets orbit which star.
    \label{fig:4plsystems_teff}
\end{figure*}

In Figure \ref{fig:4plsystems_teff}, we compare TOI-1246 to other exoplanetary systems with four confirmed planets with measured radii. We exclude Kepler-37, Kepler-48, Kepler-411, Kepler-65, and WASP-47 because they only have measured radii for 3 planets, as well as HR 8799 for visual clarity, as it hosts four very long-period planets discovered using the direct imaging method. TOI-1246 is one of only six four-planet systems with both measured masses and radii for all planets in the system. The other five systems are K2-266 \citep{rodriguez+2018}, KOI-94 \citep{weiss+2013}, K2-32 \citep{lillo-box+2020}, Kepler-223 \citep{mills+2016}, and Kepler-79 \citep{jontof-hutter+2014}. Visual inspection of this small sample shows that TOI-1246 has a unique architecture within the group, with two closely-packed planets, (at least) two planets further out, and with the most massive planet furthest from the host star. Furthermore, of this subset of four-planet systems, TOI-1246 is the brightest host star in V magnitude, making it particularly amenable for future follow-up observations. We note that all four-planet systems with measured masses for any number of planets have bright host stars. This is a consequence of the difficulty of measuring masses in multi-planet systems through the radial velocity method, which is ameliorated somewhat for bright host stars. However, TOI-1246 is brighter than 90\% of the 188 systems that host planets exhibiting TTVs\footnote{NASA Exoplanet Archive \citep{NEA}, \url{https://exoplanetarchive.ipac.caltech.edu/cgi-bin/TblView/nph-tblView?app=ExoTbls&config=PS&constraint=default_flag=1}, accessed 7 November 2021}, which have a wide range of V magnitudes (8.93 to 17.02 mag). 

\subsection{System Formation} \label{sec:atm}
Despite the prevalence of sub-Neptune exoplanets \citep{howard+2012}, their formation pathway remains unclear \citep{bean+2020}. Sub-Neptunes are believed to be mostly rocky planets (or water worlds) that, unlike super-Earths, have retained their primary atmospheres \citep{lopez+fortney2014}. The most widely discussed formation mechanism is inward movement through the protoplanetary disk (rather than in situ formation, \citealt{inamadar+schlichting2015}). However, whether this inward movement is through the migration of planetary cores (migration model) or through the drift model, where mass growth mostly takes place close-in, is not known. Regardless of the specific mechanism of inward movement, protoplanets accrete their atmospheres from the gas disk. However, the drift and migration mechanisms do predict different atmospheric compositions; the drift model predicts volatile-poor planets as pebbles are expected to lose volatiles as they migrate inwards across the snow line \citep{Ida+2019}, while the migration model predicts a variety of volatile contents across a planetary system \citep{raymond+2018}.

We estimated how much material was available to form the TOI-1246 planets assuming that the TOI-1246 planets formed in situ at their respective semi-major axes in a minimum mass solar nebula (MMSN) with a surface density profile for solids
\begin{equation}
    \Sigma = 33FZ_{rel}(\frac{a}{AU})^{-\frac{3}{2}} \rm{g cm^{-2}}
\end{equation} 
We used F = 1 and $\rm{Z_{rel} = 0.33}$ as per \citet{chiang+youdin2010}, and found that for each planet in the TOI-1246 system, less than $\rm{5\%}$ of the measured masses would be available at their measured semi-major axes. The total measured masses for all four transiting planets are predicted to be spread across ~5.5 AU in the disk. We also considered several other surface density prescriptions discussed in \citet{dai+2020}, including those described by \citet{chiang+laughlin2013} and \citet{schlinchting2014}. We find that none of these profiles provide sufficient mass interior to 0.25 AU to form the TOI-1246 planets at their measured masses. This indicates that the planets likely migrated inward in the disk to their current observed locations.

Transmission spectroscopy of sub-Neptunes can shed light on the formation mechanism(s) responsible for creating these planets. TOI-1246 is an interesting test bed for investigating sub-Neptune formation as it hosts four transiting sub-Neptunes (and potentially a fifth non-transiting planet), which share a host star, evolutionary history, and formation environments. We calculated the Transmission and Emission Spectroscopy Metrics (TSM and ESM) as per \citet{kempton+2018}, for the four-planets in the TOI-1246 system, and report these values in Table \ref{tab:planet}. These metrics are used to quantify how favourable a planet's atmosphere may be to transmission and emission spectroscopy. We find that TOI-1246 b and TOI-1246 d are the easiest targets for transmission spectroscopy in the system, with TSMs of 48.9 and 72.7 respectively.  Although none of the four-planets have $\rm{TSM > 90}$, which is recommended for high quality transmission spectra for 1 - 4 $\rm{R_\oplus}$ planets, they still present a compelling opportunity to compare the atmospheres of planetary siblings in the same system. 

There are eight other multi-planet systems with two or more planets with TSMs equal to or greater than that of TOI-1246 b, which has the second-highest TSM in the TOI-1246 system. Kepler 51 \citep{steffen+2013, masuda2014} is one such system, with two low-density `super-puff' planets that were probed using HST/WFC3 \citep{libby-roberts+2020}. The planets were found to have featureless spectra, which were interpreted as evidence for high-altitude aerosol layers. Another promising multi-planet system for transmission spectroscopy is TOI-178 \citep{leleu+2021}, which hosts six planets. JWST will observe three of these planets under GO 2319 (PI: Hooton). TOI-1246 presents another opportunity to study the atmospheres of planetary siblings in order to glean information about system formation and evolution, and thus the TOI-1246 planets are compelling targets for the upcoming JWST mission. Furthermore, this transiting multi-planet system could be a benchmark target for the Ariel mission \citep{edwards+2019}, which will study planet formation and evolution as well as TTVs as a complimentary science case.

\subsection{A Potential Fifth Planet} \label{sec:fifth}
As discussed in Section \ref{sec:fifth_in_RVs}, we find a fifth periodic signal in the RV data at a period exterior to that of TOI-1246 e. We report the results from our preferred model with a fifth non-transiting planet candidate at $\rm{P_f = 93.8 d}$. Further observations are needed to refine the orbital period and to confirm the nature of this signal, but if it is planetary, it would be a fifth planet in an already rich multi-planet system. Furthermore, if the period is actually found to be at $\rm{76.2~d}$, this fifth planet will lie very near to the 2:1 resonance with TOI-1246 e, and in turn in 4:1 resonance with TOI-1246 d. This would imply a 4:2:1 resonant chain between these three planets, similar to that between three of Jupiter's moons: Europa, Ganymede, and Io. Such a result will allow us to probe deeper into the formation history of this system. It is thought that few resonant chains remain stable after the gas disk dissipates \citep{ogihara_ida2009}, although \citet{terquem_papaloizou2007} posits that planet pairs near commensurability (rather than in strict commensurability) may survive this era of destabilization. 

The presence of an additional planet candidate in the system may also explain the need for more RV observations than predicted in order to fully model the known planet candidates \citep{he+2021}. This motivates us to continue collecting RV data for this system to determine the period and characterize the nature of this planet candidate, while also improving the accuracy of the mass measurements for the four known transiting planets.

\section{Conclusion} \label{sec:conclusion}
The main conclusions of this work are as follows:
\begin{enumerate}
    \item We confirm four transiting sub-Neptunes around K dwarf TOI-1246, and measure the masses of TOI-1246 b and TOI-1246 c to $\rm{>7\sigma}$ precision, of TOI-1246 e to $\rm{>6\sigma}$ precision, and of TOI-1246 d to $\rm{3\sigma}$ precision.
    \item We report the following masses for the four transiting planets: $M_{b} =$ \massc $M_\oplus$, $M_{c} =$ \massd $M_\oplus$, $M_{d} =$ \massb $M_\oplus$, $M_{e} =$ \masse $M_\oplus$. We note that these measurements are robust to variations in the assumed planet eccentricities ($e_i < 1$) and number of planets ($4 \leq \rm{N_{pl} \leq 6}$).
    \item We fit the \tess{} photometry in order to refine planet radii estimates, and report the following radii: \mbox{$\rm{R_{b} =~2.97 \pm 0.06~R_\oplus}$},\mbox{$\rm{ R_{c} =~2.47 \pm 0.08~R_\oplus}$}, \mbox{$\rm{R_{d} =~3.46 \pm 0.09~R_\oplus}$}, and \mbox{$\rm{R_{e} =~3.72 \pm 0.16~R_\oplus}$}. We also find no evidence of additional transiting planets in the \tess{} time series data.
    \item The four transiting planets have relatively low bulk densities ($\rm{0.70 - 3.21~g/cm^3}$), indicating that these planets have volatile H/He envelopes, and we predict that the four planets have quite varied envelope mass fractions.
    \item We find a fifth periodic signal in the RV data, which may correspond to a planet on an orbit exterior to that of TOI-1246 e. We suggest a candidate orbital period ($\rm{93.8~d}$), and emphasize the need for additional RV observations to determine the period and characterize the nature of this signal. We include this fifth planet candidate in our RV fits and report a minimum mass of $Msin(i) = 25.6 \pm 3.6~\rm{M_\oplus}$ (which is also a $7\sigma$ mass measurement), but we note that the mass depends on the true orbital period.
    \item We find that TOI-1246 d and TOI-1246 e exhibit TTVs, due to their nearness to the 2:1 mean motion resonance.
    \item We use the SPOCK framework \citep{Tamayo+2020} to investigate the stability of this system, and find that additional un-detected planets in-between the known transiting planets would destabilize the system. Adding the fifth planet candidate exterior to TOI-1246 e does not decrease the system stability.
    \item We consider the system architecture of the TOI-1246 system in the context of other high-multiplicity systems, and find that few systems have similar patterns in period spacing. 
\end{enumerate}
TOI-1246 is a rich multi-planet system that hosts at least four sub-Neptune planets with a diversity of masses and potential compositions. Additional data will allow us to more fully characterize the fifth non-transiting planet candidate, continue to investigate TTVs, and characterize the atmospheres of these planetary siblings in the context of their host star. 

\acknowledgements
Some of the data presented herein were obtained at the W. M. Keck Observatory, which is operated as a scientific partnership among the California Institute of Technology, the University of California and the National Aeronautics and Space Administration. The Observatory was made possible by the generous financial support of the W. M. Keck Foundation. We are grateful to Natalie Batalha for her valuable contributions to the TESS-Keck Survey and for her feedback on this project.

We thank the time assignment committees of the University of California, the California Institute of Technology, NASA, and the University of Hawaii for supporting the TESS-Keck Survey with observing time at Keck Observatory.  We thank NASA for funding associated with our Key Strategic Mission Support project.  We gratefully acknowledge the efforts and dedication of the Keck Observatory staff for support of HIRES and remote observing.  We recognize and acknowledge the very significant cultural role and reverence that the summit of Maunakea has within the indigenous Hawaiian community. We are deeply grateful to have the opportunity to conduct observations from this mountain.  We thank Ken and Gloria Levy, who supported the construction of the Levy Spectrometer on the Automated Planet Finder. We thank the University of California and Google for supporting Lick Observatory and the UCO staff for their dedicated work scheduling and operating the telescopes of Lick Observatory.  This paper is based on data collected by the TESS mission. Funding for the TESS mission is provided by the NASA Explorer Program.

This paper includes data collected by the \tess{} mission. Funding for the TESS mission is provided by NASA's Science Mission Directorate. We acknowledge the use of public TOI Release data from pipelines at the \tess{} Science Office and at the \tess{} Science Processing Operations Center. Resources supporting this work were provided by the NASA High-End Computing (HEC) Program through the NASA Advanced Supercomputing (NAS) Division at Ames Research Center for the production of the SPOC data products. This research has made use of the Exoplanet Follow-up Observation Program website, which is operated by the California Institute of Technology, under contract with the National Aeronautics and Space Administration under the Exoplanet Exploration Program. This research has also made use of the NASA Exoplanet Archive, which is operated by the California Institute of Technology, under contract with the National Aeronautics and Space Administration under the Exoplanet Exploration Program.

This work also makes use of observations from the LCOGT network. Part of the LCOGT telescope time was granted by NOIRLab through the Mid-Scale Innovations Program (MSIP). MSIP is funded by NSF.

This work has made use of data from the European Space Agency (ESA) mission {\it Gaia} (\url{https://www.cosmos.esa.int/gaia}), processed by the {\it Gaia} Data Processing and Analysis Consortium (DPAC, \url{https://www.cosmos.esa.int/web/gaia/dpac/consortium}). Funding for the DPAC has been provided by national institutions, in particular the institutions participating in the {\it Gaia} Multilateral Agreement.

This research was made possible through the use of the AAVSO Photometric All-Sky Survey (APASS), funded by the Robert Martin Ayers Sciences Fund and NSF AST-1412587. This publication also makes use of data products from the Two Micron All Sky Survey, which is a joint project of the University of Massachusetts and the Infrared Processing and Analysis Center/California Institute of Technology, funded by the National Aeronautics and Space Administration and the National Science Foundation. 

This publication makes use of data products from the Wide-field Infrared Survey Explorer, which is a joint project of the University of California, Los Angeles, and the Jet Propulsion Laboratory/California Institute of Technology, funded by the National Aeronautics and Space Administration.

This work is based on observations made with the Italian Telescopio Nazionale Galileo (TNG) operated on the island of La Palma by the Fundaci\'on Galileo Galilei of the INAF (Istituto Nazionale di Astrofisica) at the Spanish Observatorio del Roque de los Muchachos of the Instituto de Astrofisica de Canarias  under programs CAT19A\_162, ITP19\_1 and A41TAC\_49. Part of this work is done under the framework of the KESPRINT collaboration (\url{http://kesprint.science}). KESPRINT is an international consortium devoted to the characterization and research of exoplanets discovered with space-based missions. This work is partly financed by the Spanish Ministry of Economics and Competitiveness through grants PGC2018-098153-B-C31.

Some of the observations in the paper made use of the High-Resolution Imaging instrument ‘Alopeke obtained under Gemini LLP Proposal Number: GN/S-2021A-LP-105. ‘Alopeke was funded by the NASA Exoplanet Exploration Program and built at the NASA Ames Research Center by Steve B. Howell, Nic Scott, Elliott P. Horch, and Emmett Quigley. 'Alopeke was mounted on the Gemini North telescope of the international Gemini Observatory, a program of NSF’s NOIR Lab, which is managed by the Association of Universities for Research in Astronomy (AURA) under a cooperative agreement with the National Science Foundation on behalf of the Gemini partnership: the National Science Foundation (United States), National Research Council (Canada), Agencia Nacional de Investigación y Desarrollo (Chile), Ministerio de Ciencia, Tecnología e Innovación (Argentina), Ministério da Ciência, Tecnologia, Inovações e Comunicações (Brazil), and Korea Astronomy and Space Science Institute (Republic of Korea).

C.D. gratefully acknowledges support from the David \& Lucile Packard Foundation and the Alfred P. Sloan Foundation. A.A.B., B.S.S. and I.A.S. acknowledge the support of Ministry of Science and Higher Education of the Russian Federation under the grant 075-15-2020-780(N13.1902.21.0039). JK gratefully acknowledges the support of the Swedish National Space Agency (SNSA; DNR 2020-00104). A.W.M. is supported by the NSF Graduate Research Fellowship grant no. DGE 1752814. J.M.A.M. is supported by the National Science Foundation Graduate Research Fellowship Program under Grant No. DGE-1842400. J.M.A.M. acknowledges the LSSTC Data Science Fellowship Program, which is funded by LSSTC, NSF Cybertraining Grant No. 1829740, the Brinson Foundation, and the Moore Foundation; his participation in the program has benefited this work. C.K.H.~acknowledges support from the National Science Foundation Graduate Research Fellowship Program under Grant No.~DGE 2146752. M.R. is supported by the National Science Foundation Graduate Research Fellowship Program under Grant Number DGE-1752134. R.A.R. is supported by an NSF Graduate Research Fellowship, grant No. DGE 1745301. P. D. is supported by a National Science Foundation (NSF) Astronomy and Astrophysics Postdoctoral Fellowship under award AST-1903811. R.L. acknowledges financial support from the Centre of Excellence \arcsec Severo Ochoa\arcsec award to the Instituto de Astrofísica de Andalucía (SEV-2017-0709). D.H. acknowledges support from the Alfred P. Sloan Foundation and the National Aeronautics and Space Administration (80NSSC20K0593, 80NSSC21K0652). TM acknowledges financial support from the Spanish Ministry of Science and Innovation (MICINN) through the Spanish State Research Agency, under the Severo Ochoa Program 2020-2023 (CEX2019-000920-S). K.W.F.L. acknowledge support by DFG grants RA714/14-1 within the DFG Schwerpunkt SPP 1992, “Exploring the Diversity of Extrasolar Planets”.

This research made use of Lightkurve, a Python package for Kepler and TESS data analysis (Lightkurve Collaboration, 2018). This research also made use of Astropy,\footnote{\url{http://www.astropy.org}} a community-developed core Python package for Astronomy \citep{exoplanet:astropy13, exoplanet:astropy18}. This research made use of \textit{exoplanet} and its dependencies \citep{exoplanet:agol20, exoplanet:astropy13, exoplanet:astropy18,exoplanet:exoplanet, exoplanet:kipping13, exoplanet:luger18, exoplanet:pymc3,exoplanet:theano, exoplanet:vaneylen19}. This work made use of \texttt{tpfplotter} by J. Lillo-Box (publicly available at www.github.com/jlillo/tpfplotter), which also made use of the python packages Astropy, Lightkurve, Matplotlib and NumPy. 

\facilities{TESS, Keck:I (HIRES), TNG (HARPS-N), Keck:II (NIRC2), Gemini:Gillett (`Alopeke), LCOGT, Dragonfly Telephoto Array, TRES (Tillinghast), CAHA (Astralux), Exoplanet Archive, Gaia}

\software{Astropy \citep{astropy:2013, astropy:2018}, Astroquery \citep{astroquery}, \texttt{batman} \citep{batman},  emcee \citep{emcee},  SpecMatch \citep{specmatch-synth}, \textit{exoplanet} \citep{exoplanet:agol20}, Lightkurve \citep{lightkurve}, \texttt{RadVel} \citep{radvel}, SPOCK \citep{Tamayo+2020}, Transit Least Squares \citep{hippke+2019}, \texttt{kiauhoku} \citep{claytor+2020}, Matplotlib \citep{matplotlib}, NumPy \citep{numpy}, \texttt{astrasens} \citep{lillobox+2012, lillo-box14b}, AstroImageJ \citep{Collins:2017}, TAPIR \citep{Jensen:2013}, celerite \citep{Foreman-Mackey2017}}

\bibliography{main}

\begin{longtable*}{cccccc}
\caption{RVs and spectral activity indicators measured from Keck/HIRES and TNG/HARPS-N}
\\
\hline
\hline
Time ($\rm{BJD_{TDB}}$)& RV (m/s)& RV Unc. (m/s)& S index&S Unc.&Instrument \\
\hline \\
\endfirsthead

\caption{\textbf{Continued} \\ RVs and spectral activity indicators measured from Keck/HIRES and TNG/HARPS-N}
\label{tab:rvdata} \\
\hline
\hline
Time ($\rm{BJD_{TDB}}$)& RV (m/s)& RV Unc. (m/s)& S index&S Unc. & Instrument \\
\hline
\endhead
2458917.0623 &    4.6671 &          1.7863 &   0.1336 &         0.0010 &      HIRES \\
2458918.0658 &    8.9729 &          1.6653 &   0.1505 &         0.0010 &      HIRES \\
2458919.0551 &    0.5704 &          1.5308 &   0.1527 &         0.0010 &      HIRES \\
2458995.8758 &    5.9488 &          1.9148 &   0.1480 &         0.0010 &      HIRES \\
2458999.8927 &   15.7512 &          1.7060 &   0.1524 &         0.0010 &      HIRES \\
2459002.9282 &    3.3317 &          1.6117 &   0.1468 &         0.0010 &      HIRES \\
2459003.8913 &    0.6774 &          1.5635 &   0.1430 &         0.0010 &      HIRES \\
2459006.8841 &   -3.7627 &          1.6187 &   0.1448 &         0.0010 &      HIRES \\
2459013.8732 &   -3.1443 &          1.6608 &   0.1465 &         0.0010 &      HIRES \\
2459016.8749 &   -4.0037 &          1.8443 &   0.1479 &         0.0010 &      HIRES \\
2459024.8693 &   -2.0563 &          1.6039 &   0.1303 &         0.0010 &      HIRES \\
2459027.8384 &   -3.0241 &          1.3825 &   0.1461 &         0.0010 &      HIRES \\
2459030.8929 &    3.0806 &          1.6265 &   0.1492 &         0.0010 &      HIRES \\
2459034.8557 &    6.1822 &          1.6041 &   0.1492 &         0.0010 &      HIRES \\
2459036.7920 &    4.8777 &          1.4543 &   0.1466 &         0.0010 &      HIRES \\
2459038.8405 &    0.2686 &          1.5753 &   0.1512 &         0.0010 &      HIRES \\
2459069.0098 &    5.4160 &          2.5534 &   0.1240 &         0.0010 &      HIRES \\
2459071.9366 &    5.9474 &          1.7575 &   0.1447 &         0.0010 &      HIRES \\
2459072.8800 &    7.4601 &          1.7438 &   0.1478 &         0.0010 &      HIRES \\
2459077.8834 &   11.5812 &          1.6499 &   0.1402 &         0.0010 &      HIRES \\
2459086.8748 &   -9.1120 &          2.2852 &   0.0948 &         0.0010 &      HIRES \\
2459089.8754 &    0.7513 &          1.6082 &   0.1397 &         0.0010 &      HIRES \\
2459090.8077 &   -0.0182 &          1.6396 &   0.1532 &         0.0010 &      HIRES \\
2459091.8104 &  -10.2708 &          1.6802 &   0.1496 &         0.0010 &      HIRES \\
2459092.8046 &   -6.7398 &          1.5881 &   0.1521 &         0.0010 &      HIRES \\
2459094.7903 &    3.0606 &          1.8001 &   0.1478 &         0.0010 &      HIRES \\
2459097.8740 &   -3.7399 &          1.8511 &   0.1466 &         0.0010 &      HIRES \\
2459101.7734 &    0.7005 &          1.5959 &   0.1514 &         0.0010 &      HIRES \\
2459114.7515 &   -2.8354 &          1.6032 &   0.1556 &         0.0010 &      HIRES \\
2459115.7854 &    0.6398 &          1.5584 &   0.1394 &         0.0010 &      HIRES \\
2459117.7538 &   -1.5928 &          1.5953 &   0.1579 &         0.0010 &      HIRES \\
2459118.7697 &    1.9755 &          1.5657 &   0.1563 &         0.0010 &      HIRES \\
2459119.7620 &   -3.0504 &          1.9919 &   0.1509 &         0.0010 &      HIRES \\
2459120.7379 &   -0.0511 &          1.7190 &   0.1466 &         0.0010 &      HIRES \\
2459121.7300 &  -13.9248 &          1.7669 &   0.1520 &         0.0010 &      HIRES \\
2459122.7450 &   -7.8657 &          1.5958 &   0.1537 &         0.0010 &      HIRES \\
2459123.7347 &    0.1256 &          1.6022 &   0.1497 &         0.0010 &      HIRES \\
2459153.7107 &    7.9963 &          1.6877 &   0.1355 &         0.0010 &      HIRES \\
2459269.1356 &  -12.3450 &          1.7423 &   0.1277 &         0.0010 &      HIRES \\
2459296.1010 &    6.7841 &          1.5442 &   0.1659 &         0.0010 &      HIRES \\
2459297.0478 &    3.1049 &          1.6319 &   0.1519 &         0.0010 &      HIRES \\
2459300.0061 &   -4.1661 &          1.9683 &   0.1603 &         0.0010 &      HIRES \\
2459314.0783 &   -1.3336 &          1.4255 &   0.1637 &         0.0010 &      HIRES \\
2459353.8600 &    6.1707 &          1.6943 &   0.1601 &         0.0010 &      HIRES \\
2459354.9439 &    6.4694 &          1.6786 &   0.1557 &         0.0010 &      HIRES \\
2459358.9110 &   -9.5293 &          1.6953 &   0.1541 &         0.0010 &      HIRES \\
2459361.9457 &    8.0640 &          1.4567 &   0.1597 &         0.0010 &      HIRES \\
2459373.8191 &    6.6260 &          1.6829 &   0.1592 &         0.0010 &      HIRES \\
2459377.0715 &   -5.2521 &          1.5642 &   0.1455 &         0.0010 &      HIRES \\
2459377.8308 &    5.2305 &          1.6516 &   0.1611 &         0.0010 &      HIRES \\
2459378.8909 &    5.2943 &          1.4859 &   0.1599 &         0.0010 &      HIRES \\
2459379.9073 &   -6.2703 &          1.6647 &   0.1617 &         0.0010 &      HIRES \\
2459383.0151 &    0.9678 &          1.6524 &   0.1595 &         0.0010 &      HIRES \\
2459383.9827 &    2.8946 &          1.6310 &   0.1536 &         0.0010 &      HIRES \\
2459385.8374 &   -8.5394 &          1.6133 &   0.1542 &         0.0010 &      HIRES \\
2459388.0531 &   -3.7768 &          1.6463 &   0.1522 &         0.0010 &      HIRES \\
2459388.9024 &  -10.0083 &          1.6023 &   0.1649 &         0.0010 &      HIRES \\
2459389.8761 &  -12.9257 &          1.5479 &   0.1585 &         0.0010 &      HIRES \\
2459395.9321 &   -4.7191 &          1.5397 &   0.1573 &         0.0010 &      HIRES \\
2459399.8640 &   -9.5533 &          1.7150 &   0.1601 &         0.0010 &      HIRES \\
2459404.9747 &   -2.2866 &          1.4522 &   0.1535 &         0.0010 &      HIRES \\
2459406.8859 &   -9.5983 &          1.6463 &   0.1559 &         0.0010 &      HIRES \\
2459407.9203 &   -0.8518 &          1.8118 &   0.0000 &         0.0010 &      HIRES \\
2459408.9688 &    4.6775 &          1.5936 &   0.1521 &         0.0010 &      HIRES \\
2459409.9706 &   -1.8515 &          1.7468 &   0.1539 &         0.0010 &      HIRES \\
2459412.9687 &    3.5893 &          1.5592 &   0.1542 &         0.0010 &      HIRES \\
2459413.9763 &    5.0963 &          2.2444 &   0.1394 &         0.0010 &      HIRES \\
2459414.9853 &    4.5878 &          2.6134 &   0.1410 &         0.0010 &      HIRES \\
2459435.7827 &    3.1791 &          1.6335 &   0.1554 &         0.0010 &      HIRES \\
2459441.8991 &    2.8895 &          1.6247 &   0.1604 &         0.0010 &      HIRES \\
2459444.9312 &    6.0879 &          1.6149 &   0.1580 &         0.0010 &      HIRES \\
2459448.8678 &   10.8938 &          1.9052 &   0.1563 &         0.0010 &      HIRES \\
2459449.8081 &    7.5028 &          1.6172 &   0.1598 &         0.0010 &      HIRES \\
2459450.8254 &    1.0285 &          1.7628 &   0.1448 &         0.0010 &      HIRES \\
2459451.8711 &    9.3126 &          1.6446 &   0.1652 &         0.0010 &      HIRES \\
2459452.8133 &    6.6857 &          1.5619 &   0.1581 &         0.0010 &      HIRES \\
2459455.7976 &    7.1597 &          1.5236 &   0.1641 &         0.0010 &      HIRES \\
2459456.8270 &    3.4921 &          1.6717 &   0.1673 &         0.0010 &      HIRES \\
2459469.7871 &   -6.1427 &          1.6299 &   0.1604 &         0.0010 &      HIRES \\
2459470.7829 &   -5.3842 &          1.6701 &   0.1623 &         0.0010 &      HIRES \\
2459472.7681 &   -3.8466 &          1.6613 &   0.1483 &         0.0010 &      HIRES \\
2459475.7760 &  -17.8983 &          1.8216 &   0.1581 &         0.0010 &      HIRES \\
2459476.7713 &  -10.8095 &          1.6625 &   0.1597 &         0.0010 &      HIRES \\
2459478.7999 &    0.7152 &          1.7799 &   0.1648 &         0.0010 &      HIRES \\
2459482.7845 &   -2.5158 &          1.5947 &   0.1629 &         0.0010 &      HIRES \\
2459483.7797 &    0.8174 &          1.6887 &   0.1588 &         0.0010 &      HIRES \\
2459484.7623 &   -1.0387 &          1.6102 &   0.1554 &         0.0010 &      HIRES \\
2459489.7755 &   -3.6559 &          1.7329 &   0.1475 &         0.0010 &      HIRES \\
2459497.7223 &   -5.6037 &          2.1168 &   0.1585 &         0.0010 &      HIRES \\
2459498.7418 &   -5.0932 &          1.9443 &   0.1624 &         0.0010 &      HIRES \\
2459502.7756 &    5.0014 &          1.7797 &   0.1572 &         0.0010 &      HIRES \\
2459503.7694 &    4.7857 &          1.8185 &   0.1497 &         0.0010 &      HIRES \\
2459504.7741 &  -12.2155 &          1.8252 &   0.1666 &         0.0010 &      HIRES \\
2459506.7250 &   -3.0769 &          1.8514 &   0.1676 &         0.0010 &      HIRES \\
2459508.7248 &    3.8190 &          1.7255 &   0.1642 &         0.0010 &      HIRES \\
2459513.7437 &    1.5101 &          1.7032 &   0.1223 &         0.0010 &      HIRES \\
2459516.7123 &    3.2775 &          1.8132 &   0.1641 &         0.0010 &      HIRES \\
2459622.1271 &   -0.5991 &          1.8356 &   0.1579 &         0.0010 &      HIRES \\
2459626.1265 &   -1.3294 &          1.7349 &   0.1516 &         0.0010 &      HIRES \\
2459632.1152 &    7.7221 &          1.6469 &   0.1643 &         0.0010 &      HIRES \\
2458896.7475 &    4.5793 &          2.1153 &   0.1760 &         0.0126 &    HARPS-N \\
2458897.7587 &    0.6066 &          1.7459 &   0.1624 &         0.0093 &    HARPS-N \\
2458898.7442 &   -0.3685 &          1.6674 &   0.1663 &         0.0101 &    HARPS-N \\
2458904.7360 &    3.8349 &          1.6534 &   0.1397 &         0.0096 &    HARPS-N \\
2458905.7339 &   10.4724 &          1.5069 &   0.1617 &         0.0057 &    HARPS-N \\
2458925.7178 &    5.6357 &          1.7037 &   0.1601 &         0.0076 &    HARPS-N \\
2458926.7187 &    9.6815 &          1.3208 &   0.1559 &         0.0056 &    HARPS-N \\
2458929.7005 &    9.5106 &          1.1660 &   0.1508 &         0.0047 &    HARPS-N \\
2459000.5461 &   11.8829 &          1.3999 &   0.1638 &         0.0074 &    HARPS-N \\
2459000.6312 &   14.3818 &          1.4844 &   0.1728 &         0.0084 &    HARPS-N \\
2459002.5486 &   -0.3836 &          1.0787 &   0.1681 &         0.0044 &    HARPS-N \\
2459002.6361 &   -2.5666 &          1.2797 &   0.1719 &         0.0052 &    HARPS-N \\
2459012.5520 &   -0.9495 &          1.3709 &   0.1633 &         0.0067 &    HARPS-N \\
2459012.6414 &   -1.3279 &          1.1019 &   0.1626 &         0.0050 &    HARPS-N \\
2459014.6408 &   -9.3870 &          1.4438 &   0.1606 &         0.0066 &    HARPS-N \\
2459087.3945 &   -6.0197 &          1.7265 &   0.1731 &         0.0105 &    HARPS-N \\
2459089.3867 &   -3.9607 &          2.5947 &   0.1394 &         0.0192 &    HARPS-N \\
2459248.7604 &    5.5201 &          1.3695 &   0.1714 &         0.0067 &    HARPS-N \\
2459268.7188 &   -7.4858 &          2.4539 &   0.1824 &         0.0180 &    HARPS-N \\
2459309.6076 &    1.4668 &          1.8518 &   0.1718 &         0.0091 &    HARPS-N \\
2459310.6164 &   -2.3908 &          2.2873 &   0.1684 &         0.0151 &    HARPS-N \\
2459353.6993 &    5.6099 &          2.0655 &   0.1632 &         0.0166 &    HARPS-N \\
2459353.7169 &    3.8382 &          1.9814 &   0.1673 &         0.0158 &    HARPS-N \\
2459370.5405 &    2.1340 &          2.4140 &   0.1647 &         0.0169 &    HARPS-N \\
2459393.6381 &  -11.8390 &          1.3336 &   0.1672 &         0.0070 &    HARPS-N \\
2459394.5373 &  -10.6052 &          1.5538 &   0.1824 &         0.0095 &    HARPS-N \\
2459410.4726 &    4.9166 &          2.1437 &   0.1900 &         0.0133 &    HARPS-N \\
2459450.4350 &    7.3945 &          1.3333 &   0.1726 &         0.0064 &    HARPS-N \\
\hline
\end{longtable*}

\end{document}